\documentclass[preprint,aps,tightenlines,showpacs]{revtex4}
\usepackage[dvips]{graphicx}
\usepackage{dcolumn}
\usepackage{epsfig}

\begin{document}
\title{On the Determination of $\bar{d}/\bar{u}$ Ratios from Proton-Proton and
 Proton-Deuteron Drell-Yan processes} 
\author{T.-S. H. Lee}
\affiliation{Physics Division, Argonne National Laboratory,
Argonne, Illinois 60439, USA}

\begin{abstract}
By using the convolution formula derived within the framework of
relativistic quantum mechanics,
we have examined the  Fermi Motion effects 
on the ratios
$R_{pd/pp}=\sigma^{pd}/(2\sigma^{pp})$ between the proton-proton ($p$-$p$) and
proton-deuteron ($p$-$d$) Drell-Yan cross sections. 
We have found that  in the small $x_2 < 0.3$ region,
the Fermi Motion effect is less than $1\%$ and
our results for the ratios $R_{pd/pp}=\sigma^{pd}/(2\sigma^{pp})$
 agree well with the data at 800 GeV.
In the large
 Bjorken $x_2 > $ about 0.4 region,
the  $p$-$d$ Drell-Yan cross sections
can be influenced strongly by the Fermi motion effect.
At $120$ GeV
the predicted Fermi Motion effect can enhance the ratios $R_{pd/pp}$ by
about 20 $\%$ at $x_2 \sim 0.6$ and about a factor of $2.5$ at
$x_2 \rightarrow 1.0$.
Our results suggest that the Fermi Motion effect,
along with other possible nuclear effects, must be included,
in extracting the $\bar{d}/\bar{u}$ ratios in the proton
 from the experiments
on $p$-$p$ and $p$-$d$ Drell-Yan processes at large $x_2$.
\end{abstract}

\maketitle

\section{Introduction}
Since the asymmetry between the anti-up($\bar{u}$) quark            
and anti-down ($\bar{d}$) quark distributions in the proton was revealed
by the New Muon Collaboration\cite{nmc} (NMC),
a series of
experiments\cite{peng1992,peng1998,peng1998a,peng2001} on the
di-muons ($\mu^+\mu^-)$ production from the Drell-Yan\cite{dy} (DY) processes in
$p+p$ and $p+d$ collisions had been performed at 
Fermi National Accelerator Laboratory (Fermi Laboratory).
 The objective was to extract
the $\bar{d}/\bar{u}$
ratio of the parton distribution functions (PDF) in the proton.
The information from these experiments and
 the measurements\cite{nmc,na51,hermes}
 of deep inelastic 
scattering (DIS) of leptons from the nucleon have confirmed the NMC's finding,
 $\bar{d}/\bar{u} > 1 $, 
 in the region of low Bjorkin $x \leq $ about 0.3.  

The ratio $\bar{d}/\bar{u} >1 $ 
signals the nonperturbative nature of the sea of the proton.
Its dynamical origins have been
investigated\cite{thom83,thom98,kuma98,spet97,kuma91,hwan93,szcz94,koep96,
niko99,albe00,eich93,szcz96,poby99,doro93,meln99}
 rather extensively. Precise experimental determination
of $\bar{d}/\bar{u}$ for higher $x > $ 0.3 is needed to distinguish more 
decisively these models and to develop a deeper understanding of 
the sea of the proton.
This information will soon become available from a forthcoming 
experiment\cite{peng2011} at Fermi Laboratory.

It is instructive to describe here how the $p$-$d$ DY data were analyzed.
In the analysis of the data of Ref.\cite{peng2001}, 
the leading-order DY cross section of the $p+N$ collision with 
$N= p$ (proton), $n$ (neutron) is written s
\begin{eqnarray}
\frac{d\sigma^{pN}}{dx_1 dx_2} =
\frac{4\pi \alpha^2}{9 M^2} \sum_{q}e^2_q[f^q_p(x_1) f^{\bar{q}}_N(x_2)
+f^{\bar{q}}_p(x_1) f^{\bar{q}}_N(x_2)] \,,
\label{eq:dy-exp}
\end{eqnarray}
where the sum is over all quark flavors, $\alpha=1/137$, $e_q$ is the quark
charge, $f^q_N(x)$ is the parton distribution of parton $q$ in hadron
$N$, and $M$ is the virtual photon or di-lepton mass.
Here $x_1$ and $x_2$ are the Bjorken-$x$ of partons from the beam ($p$)
and target ($N$), respectively.
The DY cross section for $p+d$ is simply taken as
\begin{eqnarray}
\frac{d\sigma^{pd}}{dx_1 dx_2} = \frac{d\sigma^{pp}}{dx_1 dx_2}
+\frac{\sigma^{pn}}{dx_1 dx_2}\,.
\label{eq:dy-pd-exp}
\end{eqnarray}
Clearly the effects due to the  nucleon Fermi Motion 
in the deuteron are not included in Eq.(\ref{eq:dy-pd-exp}).
The use of Eq.(\ref{eq:dy-pd-exp}) seems valid to a very large extent
in the small $x_2$ region since the
available data on nuclei indicate that  DY cross section per nucleon is rather independent
of the nuclear mass number $A$.
However it is not clear  whether it
can be used to extract the $\bar{d}/\bar{u}$ ratios in the proton from
the upcoming experiments at large $x_2$.

In the DIS studies\cite{disreview-11},
 it is well recognized that the
nuclear effects
must be considered in extracting the parton distributions of the nucleon
from the data. In particular it is mandatory to include nucleon Fermi motion effects
by using various forms of convolution formula to express the DIS cross sections in terms of
nucleon momentum distributions and parton distribution functions.
In contrast, there exists very limited efforts to include
 the similar nuclear effects in analyzing the DY data on
the deuteron\cite{peng1992,peng1998,peng1998a,peng2001}
and nuclei\cite{miller84,dynucl-1990,dynucl-1999,dypi-1989,holt2005}.
As a step to improve the situation, it is necessary to
 investigate  under
what assumptions 
 Eqs.(\ref{eq:dy-exp})-(\ref{eq:dy-pd-exp})
can be
derived from a formulation within which the effects due to
the internal motions
of partons in the nucleon and nucleons in the deuteron can be 
defined rigorously and consistently. This is the main objective of this work.

We will develop convolution formula for calculating the $p$-$d$ DY cross sections
within 
the relativistic quantum mechanics proposed by Dirac\cite{dirac},
as reviewed by Keister and Polyzou\cite{kp-book}.
The same theoretical framework was taken in the studies of
 electron-deuteron 
scattering\cite{coester-a,polyzou-a}, electron-$^3He$ scattering\cite{coester-b}, 
and DIS on deuteron\cite{achl09}. 
There exists convolution formula for DY processes on nuclei 
, as given, for example,  in
Ref.\cite{miller84}. However it is not clear how
those formula  can be related 
 to the relativistic formulation considered in our approach.
To explain clearly the content of our approach,
we thus will give a rather elementary derivation of our formula with
all approximations specified explicitly.
We will  apply our formula to analyze the available data
 at 800 GeV\cite{peng2001} and make predictions
for the forthcoming experiment\cite{peng2011}.

In section II, we start with the general
covariant form of the DY cross section and indicate the procedures needed to
obtain  the well known
 $q\bar{q} \rightarrow \mu^+\mu^-$ cross section $\sigma^{q\bar{q}}$.
The same procedures are used in section III
to derive formula for
calculating the $p+N$  DY cross sections from
$\sigma^{q\bar{q}}$ and the properly defined  
parton distribution functions $f^{q}_N$ of the nucleon.
In section IV, we use the impulse approximation to derive
the convolution formula for calculating $p+d$ DY cross sections from
$\sigma^{q\bar{q}}$, $f^{q}_N$, and the nucleon momentum distribution
$\rho_{p_d}(\vec{p}_N)$ 
of a fast moving  deuteron with momentum $p_d$. 

In section V, we describe the procedures for applying the developed formula
to perform  numerical calculations
of $p+p$ and $p+d$ DY cross sections
using the available parton 
distributions\cite{lai97,mart96,gluc95,plot95,ceteq5m} and
realistic deuteron wavefunctions\cite{v18,nij,cdbonn,paris}. 
The results are presented in section VI.
We will compare our results with the
available data at 800 GeV\cite{peng2001} and make predictions at 120 GeV
for analyzing the forthcoming experiment\cite{peng2011}.
A summary and discussions on necessary future improvements are given in
section VII.

\section{Covariant formula for Drell-Yan Cross sections}
The formula presented in this section are derived from using
the  Bjorken-Drell\cite{bd-book} conventions for the
Dirac matrices and the field operators for Fermions and photons.
To use the formula of Relativistic Quantum Mechanics given in Ref.\cite{kp-book},
the plane-wave state
$|\vec{k}>$ is normalized as
$<\vec{k}|\vec{k}^{\,'}>=\delta(\vec{k}-\vec{k}^{\,'})$ and the bound
states $|\Phi_\alpha>$ of composite particles, nucleons or nuclei,
are normalized as $<\Phi_\alpha|\Phi_\beta> =\delta_{\alpha,\beta}$.
To simplify the presentation, spin indices are
suppressed;
i.e. $|\vec{k}_a>$ represents $|\vec{k}_a,\lambda_a>$ for a particle
$a$ with helicity $\lambda_a$.
Thus the formula presented here are only for the spin averaged cross sections
which are the focus of this paper.

\begin{figure}[t]
\begin{center}
\includegraphics[clip,width=0.5\textwidth]{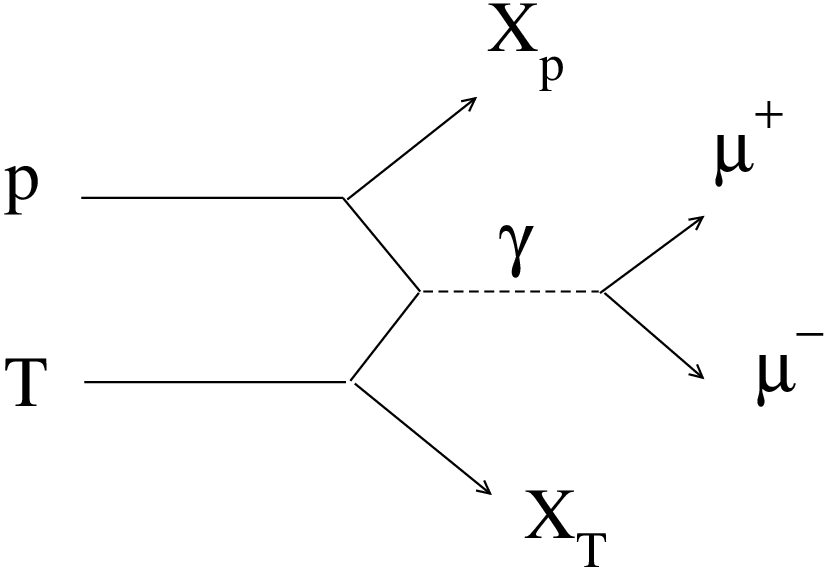}
\caption{
Drell-Yan process.}
 \label{fig:dy-mech}
\end{center}
\end{figure}
We consider  the di-muons production from the DY processes of  
 hadron ($h$) - hadron ($T$) collisions:
\begin{eqnarray}
h(p_h) + T(p_T) \rightarrow \mu^+(k_+) + \mu^-(k_{-})+X_h(p_{x_h}) 
+ X_T(p_{x_T}) \,,
\label{eq:dy-process}
\end{eqnarray}
 where $X_h$ and $X_T$ are the  undetected fragments, and the four-momentum of each particle is given
within the parenthesis. 
In terms of the partonic 
$q\bar{q}\rightarrow \gamma \rightarrow \mu^++\mu^-$ mechanism,
illustrated in Fig.\ref{fig:dy-mech}, 
the covariant form of the di-muons production cross section
 can be written as 
\begin{eqnarray}
d\sigma=\frac{(2\pi)^4}{4[(p_h\cdot p_T)^2-m^2_hm^2_T]^{1/2}}
\frac{1}{(2\pi)^6}\frac{d\vec{k}_+}{2E_+}\frac{d\vec{k}_-}{2E_-}
\frac{1}{q^4}f^{\mu\nu}(k_+,k_-)F_{\mu\nu}(p_h,p_T,q) \,, 
\label{eq:dy-crst}
\end{eqnarray}
where $m_h$ and $m_T$ are the masses for $h$ and $T$, respectively,
$ E_{\pm}=[\vec{k}^{\,2}_{\pm} + m^2_\mu]^{1/2}$ are the energies of 
muons $\mu^{\pm}$, and 
$ q = k_+ + k_- $ is the momentum of the virtual photon.
The leptonic tensor is defined by
\begin{eqnarray}
f^{\mu\nu}(k_+,k_-) =  (2\pi)^6 (2E_+)(2E_-)
<\vec{k}_+\vec{k}_-|j^\mu(0)|0><0|j^\nu(0)|\vec{k}_+\vec{k}_->\,.
\label{eq:f-lepton}
\end{eqnarray}
Here 
the leptonic current  is
\begin{eqnarray}
j^\mu(x)=e\bar{\psi}_\mu(x)\gamma^\mu \psi_\mu(x)\,,
\label{eq:c-lepton}
\end{eqnarray}
where $\psi_\mu(x)$ is the field operator for  muon, and
$e=\sqrt{4\pi\alpha}$ with $\alpha=1/137$.
By using the definitions Eqs.(\ref{eq:f-lepton})-(\ref{eq:c-lepton}), it is straightforward to
get the following analytic form of the lepton tensor
\begin{eqnarray}
f^{\mu\nu}(k_+,k_-)
&=& -4e^2[k^\mu_+ k^\nu_- +k^\nu_+ k^\mu -g^{\mu\nu}(k_+\cdot k_- +m^2_\mu)]\,.
\label{eq:lepton-t}
\end{eqnarray}

Within the parton model, the hadronic tensor in Eq.(\ref{eq:dy-crst}) is 
determined by  
the current $J_\mu(x)$ carried by partons $q$ or $\bar{q}$
\begin{eqnarray}
F_{\mu\nu}(p_h,p_T,q)&=&
\sum_{x_h,x_T} (2\pi)^6(2E_h)(2E_T)
\int d\vec{p}_{x_h}d\vec{p}_{x_T}
\delta^4(p_h+p_T-p_{x_h}-p_{x_T}-q)
\nonumber \\
&&\times
<p_h p_T|J_\mu(0)|\vec{p}_{x_h}d\vec{p}_{x_T}>
<\vec{p}_{x_T}\vec{p}_{x_h}|J_\nu(0)|p_hp_T> \,,
\label{eq:f-hadron}
\end{eqnarray}
where 
\begin{eqnarray}
J_\mu(x)=\sum_{q}[\hat{e}_q e]\bar{\psi}_{q}(x)\gamma^\mu \psi_q(x)\,.
\label{eq:c-parton}
\end{eqnarray}
Here $\psi_q(x)$ is the field operator for a quark
$q$  with charge $\hat{e}_q e$; i.e $\hat{e}_u=\frac{2}{3}$
and $\hat{e}_d=-\frac{1}{3}$ for the up and down quarks, respectively.

 The above covariant expressions are convenient for deriving 
the formula which can express the
hadron-hadron  DY cross
sections in terms of  the elementary partonic  
$q\bar{q} \rightarrow \mu^+\mu^-$ cross sections. To get such formula,
we first show how the elementary 
$q\bar{q} \rightarrow \mu^+ \mu^-$ cross section can be derived
from 
Eq.(\ref{eq:dy-crst}) with $h=q$ and $T=\bar{q}$.
Explicitly, Eq.(\ref{eq:dy-crst}) for the
 $q(p_q)+ \bar{q}(p_{\bar{q}}) \rightarrow \mu^+(k_+)+ \mu^-(k_-)$
process is
\begin{eqnarray}
d\sigma^{q\bar{q}} =\frac{(2\pi)^4}{4[(p_q\cdot p_{\bar{q}})^2
-m^4_q]^{1/2}}
\{\frac{1}{(2\pi)^6}\frac{d\vec{k}_+}{2E_+}\frac{d\vec{k}_-}{2E_-}
\frac{1}{q^4}f^{\mu\nu}(k_+,k_-)F^{q\bar{q}}_{\mu\nu}(p_q,p_{\bar{q}},q)\}\,.
\label{eq:dy-qbarq-crst}
\end{eqnarray}
The next step is replace the intermediate states 
$|\vec{p}_{x_h} \vec{p}_{x_T}>$ by the 
the vacuum state $|0>$ in evaluating
the hadronic tensor Eq.(\ref{eq:f-hadron}). We thus have
\begin{eqnarray}
F^{q\bar{q}}_{\mu\nu}(p_q,p_{\bar{q}},q)
&=&(2\pi)^6 (2E_q)(2E_{\bar{q}})\nonumber \\
&& \times 
 <p_{\bar{q}}p_q|J_\mu(0)|0>
<0|J_\nu(0)|p_q,p_{\bar{q}}>\delta^4(p_q+p_{\bar{q}}-q)\,.
\label{eq:qq-tensor}
\end{eqnarray}
Substituting parton current Eq.(\ref{eq:c-parton}) into Eq.(\ref{eq:qq-tensor}),
the hadronic tensor $F^{q\bar{q}}_{\mu\nu}$ then
has a form which is the same as the leptonic 
tensor $f^{\mu\nu}$ defined by Eqs.(\ref{eq:f-lepton})-(\ref{eq:c-lepton})
 except that the momentum variables and the charges associated with 
the Fermion field operators 
are different. By appropriately changing the momentum variables
in Eq.(\ref{eq:lepton-t}),
we obtain
\begin{eqnarray}
F^{q\bar{q}}_{\mu\nu}(p_q,p_{\bar{q}},q) 
= -4 [\hat{e}_qe]^2[p_q^\mu p_{\bar{q}}^\nu +  p_q^\nu p_{\bar{q}}^\mu 
-g^{\mu\nu}(p_q\cdot p_{\bar{q}} +m^2_q)]\delta^4(p_q+p_{\bar{q}}-q) \,.
\label{eq:parton-t}
\end{eqnarray}
By using Eqs.(\ref{eq:lepton-t}) and (\ref{eq:parton-t}), 
Eq.(\ref{eq:dy-qbarq-crst}) for the cross sections of
$q(p_q)+ \bar{q}(p_{\bar{q}}) \rightarrow \mu^+(k_+)+ \mu^-(k_-)$
can then be written as
\begin{eqnarray}
d\sigma^{q\bar{q}}(p_q,p_{\bar{q}})
&=&\frac{(2\pi)^4}{4[(p_q\cdot p_{\bar{q}})^2-m_q^4]^{1/2}}
\frac{1}{(2\pi)^6}\frac{d\vec{k}_+}{2E_+}\frac{d\vec{k}_-}{2E_-}
\frac{1}{q^4}\delta^4(p_q+p_{\bar{q}}-q) \nonumber \\
&&\times
8[k_+\cdot p_qk_-\cdot p_{\bar{q}} + k_-\cdot p_qk_+\cdot p_{\bar{q}}
+m^2_q\frac{(k_+-k_-)^2}{2} + m^2_\mu\frac{(p_q-p_{\bar{q}})^2}{2}]\,. 
\label{eq:crst-qq}
\end{eqnarray}

It is convenient to express the $q$-$\bar{q}$ DY cross section  in terms of
the invariant function $q^2 = (p_q+p_{\bar{q}})^2 = (k_{+}+k_{-})^2$.
After some derivations and accounting for the color degrees of freedom of 
quarks, we obtain
\begin{eqnarray}
\frac{d\sigma^{q\bar{q}}(p_q,p_{\bar{q}})}{dq^2} &=&
\frac{4\pi \alpha^2}{q^2}\hat{e}_q^2\frac{1}{3N_c}
\frac{[q^2-m^2_\mu/4]^{1/2}}
{[q^2-m^2_q/4]^{1/2}} 
\delta(q^2-(p_q+p_{\bar{q}})^2)\,,
\end{eqnarray}
where $N_c$ is the number of colors. Taking $N_c=3$ and 
considering  $q^2 >> m^2_\mu $ and $q^2>> m^2_q$, we then obtain the familiar form
\begin{eqnarray}
\frac{d\sigma^{q\bar{q}}(p_q,p_{\bar{q}})}{dq^2} &=&
\frac{4\pi \alpha^2}{q^2}\hat{e}_q^2\frac{1}{9}
\delta(q^2-(p_q+p_{\bar{q}})^2)\,.
\label{eq:dy-0}
\end{eqnarray}
The above expression is identical to the commonly used expression,
as given, for example, in Ref.\cite{ceteq5m}.

In the next two sections, we will derive formula expressing
 the $p$-$N$ and $p$-$d$ DY cross sections in terms of 
$d\sigma^{q\bar{q}}(p_q,p_{\bar{q}})/dq^2$ given in Eq.(\ref{eq:dy-0}).
To simplify the presentation, we only present formula for
$q$ in the projectile $p$ and $\bar{q}$ in the target $N$ or  $d$.
The term from interchanging $q\leftrightarrow \bar{q}$ will be included only
in the final expressions for calculations.

\section{$p$-$N$ DY cross sections}

Eq.(\ref{eq:dy-crst}) for the 
$p(p_1)+ N(p_2)\rightarrow \mu^+(k_+)+\mu^-(k_-) + X_p(p_{X_p})+X_N(p_{X_N}) $
process  is
\begin{eqnarray}
d\sigma^{pN} =\frac{(2\pi)^4}{4[(p\cdot p_N)^2
-m^2_pm^2_N]^{1/2}}
\{\frac{1}{(2\pi)^6}\frac{d\vec{k}_+}{2E_+}\frac{d\vec{k}_-}{2E_-}
\frac{1}{q^4}f^{\mu\nu}(k_+,k_-)F^{pN}_{\mu\nu}(p_p,p_N,q)\} \,,
\label{eq:dy-crst-pn}
\end{eqnarray}
where the hadronic tensor, defined by  Eq.(\ref{eq:f-hadron}), is
\begin{eqnarray}
F^{pN}_{\mu\nu}(p_p,p_N,q)
&=&
(2\pi)^6 (2E_p)(2E_N)\{ \sum_{X_p,X_N} \int d\vec{p}_{X_p} d\vec{p}_{X_N}
\delta^4(p_p+p_N-p_{X_p}-p_{X_N}-q)
\nonumber \\
&&\times <p_Np_p |J_\mu(0)|p_{X_p} p_{X_N}><p_{X_N} p_{X_p}|J_\nu(0)|p_pp_N>\}
\,.
\label{eq:pN-tensor}
\end{eqnarray}
Within the parton model, the
DY cross sections are calculated from  
the matrix element
$<q\bar{q}|J_\mu(0)|0><0|J_\nu(0)|q\bar{q}>$  which is
part of the matrix element describing
the annihilation of a 
$q$ ( $\bar{q}$) from the projectile $p$ and a $\bar{q}$ ($q$) from
the target $N$ into a photon. 
To identify such matrix elements, we
insert  a complete set of $q\bar{q}$ states 
\begin{eqnarray}
1 = \int d\vec{p}_{q}d\vec{p}_{\bar{q}}\,\, |\vec{p}_{q}\vec{p}_{\bar{q}}>
<\vec{p}_{\bar{q}}\vec{p}_{q}| \nonumber
\end{eqnarray}
between $<p_Np_p|$ and $J_\mu(0)$ ($J_\nu(0)$ and $|p_pp_N>$)
 in Eq.(\ref{eq:pN-tensor})
and neglect any electromagnetic contribution from the undetected fragments $X_p$ and $X_T$. 
We then have
\begin{eqnarray}
F^{pN}_{\mu\nu}(p_p,p_N,q)
&=&
(2\pi)^6 (2E_p)(2E_N) 
\sum_{X_p,X_N} \int d\vec{p}_{X_p} d\vec{p}_{X_N}
\delta^4(p_p+p_N-p_{X_p}-p_{X_N}-q) \nonumber \\
&& \times \int d\vec{p}_{q}d\vec{p}_{\bar{q}}
\int d\vec{p}^{\,'}_{q}d\vec{p}^{\,'}_{\bar{q}}
 <\vec{p}_p|\vec{p}_q\vec{p}_{X_p}>
<\vec{p}_N|\vec{p}_{\bar{q}}\vec{p}_{X_N}>
<\vec{p}_{X_p}\vec{p}^{\,'}_q| \vec{p}_p>
<\vec{p}_{X_N}\vec{p}^{\,'}_{\bar{q}}| \vec{p}_N> \nonumber \\
&&\times<\vec{p}_{\bar{q}}\vec{p}_{q}|J_\mu(0)|0><0|J_\nu(0)|\vec{p}^{\,'}_q
\vec{p}^{\,'}_{\bar{q}}>\,.
 \label{eq:pn-1}
\end{eqnarray}

By momentum conservation, the overlap functions in the above
equation can be written as
\begin{eqnarray}
<\vec{p}_q \vec{p}_{X}|\vec{p}_p> &=& <\vec{p}_X|b_{\vec{p}_q}|\vec{p}_p>
\nonumber \\
&=&\phi_{\vec{p}_p}(\vec{p}_q ,\vec{p}_{X})
\delta(\vec{p}_p-\vec{p}_q -\vec{p}_{X}) \,, \label{eq:pn-2}
\end{eqnarray}
where $b_{\vec{p}_q}$ is the annihilation operator of quark $q$.
By using the definition  Eq.(\ref{eq:pn-2}),  
Eq.(\ref{eq:pn-1}) can be written as
\begin{eqnarray}
F^{pN}_{\mu\nu}(p_p,p_N,q)
&=&
(2\pi)^6(2E_p)(2E_N)
\sum_{X_P,X_N} \int d\vec{p}_{X_p} d\vec{p}_{X_N}
\delta^4(p_p+p_N-p_{X_p}-p_{X_N}-q)
\nonumber \\
&&\times[\int d\vec{p}_q 
|\phi_{\vec{p}_p}(\vec{p}_q ,\vec{p}_{X_p})|^2
\delta(\vec{p}_p-\vec{p}_q -\vec{p}_{X_p})]
\nonumber \\
&&\times
[\int d\vec{p}_{\bar{q}}
| \phi_{\vec{p}_N}(\vec{p}_{\bar{q}} ,\vec{p}_{X_N})|^2
\delta(\vec{p}_N-\vec{p}_{\bar{q}} -\vec{p}_{X_N})]
\nonumber \\
&&\times
<\vec{p}_{\bar{q}}\vec{p}_{q}|J_\mu(0)|0><0|J_\nu(0)|\vec{p}_q
\vec{p}_{\bar{q}}> \label{eq:pn-3}
\end{eqnarray}

 The evaluation of
the expression  Eq.(\ref{eq:pn-3}) needs rather detailed information about the
undetected fragments $X_p$ and $X_N$ because of
the dependence of 
$\delta^4(p_p+p_N-p_{X_p}-p_{X_N}-q)$ on their energies $p^0_{X_p}$ and
$p^0_{X_N}$.
To simplify the calculation, we follow the common practice to 
neglect the explicit dependence of the energies $p^0_{X_p}$ and
$p^0_{X_N}$ of the undetected fragments. This amounts to neglecting the binding
effects by
setting  $p^0_{X_p}\sim \epsilon_1$ and
$p^0_{X_N}\sim \epsilon_2$ to write  
\begin{eqnarray}
\delta^4(p_p+p_N-p_{X_p}-p_{X_N}-q)
\sim
\delta^3(\vec{p}_p+\vec{p}_N-\vec{p}_{X_p}-\vec{p}_{X_N}-\vec{q})
\delta(p^0_p+p^0_N - \epsilon_1-\epsilon_2 - q^0)\,, \nonumber \\
  \label{eq:delta-f}
\end{eqnarray}
where $\epsilon_1$ and $\epsilon_2$ are constants.

We now define
\begin{eqnarray}
F^{q}_{\vec{p}_p}(\vec{p}_q) = 
\sum_{X_p} \int d\vec{p}_{X_p}
|\phi_{\vec{p}_p}(\vec{p}_q ,\vec{p}_{X_p})|^2
\delta(\vec{p}_p-\vec{p}_q -\vec{p}_{X_p})
\label{eq:q-dis} 
\end{eqnarray}
for the projectile $p$, and 
\begin{eqnarray}
F^{\bar{q}}_{\vec{p}_N}(\vec{p}_{\bar{q}}) =
\sum_{X_N} \int d\vec{p}_{X_N}
|\phi_{\vec{p}_N}(\vec{p}_{\bar{q}} ,\vec{p}_{X_N})|^2
\delta(\vec{p}_N-\vec{p}_{\bar{q}} -\vec{p}_{X_N})
\label{eq:barq-dis}
\end{eqnarray}
for  the target $N$. These two definitions and  
the approximation Eq.(\ref{eq:delta-f}) 
 allow us to cast Eq.(\ref{eq:pn-3}) 
into the following form
\begin{eqnarray}
F^{pN}_{\mu\nu}(p_p,p_N,q)
&=&
(2\pi)^6 (2E_p)(2E_N)\sum_{q}\int d\vec{p}_q d\vec{p}_{\bar{q}}
[F^q_{p_p} (p_q) F^{\bar{q}}_{p_N} (p_{\bar{q}})]
\nonumber \\
&&\times
[<\vec{p}_{\bar{q}}\vec{p}_{q}|J_\mu(0)|0><0|J_\nu(0)|\vec{p}_q
\vec{p}_{\bar{q}}>  \nonumber \\
&&\times 
\delta^3(\vec{p}_q+\vec{p}_{\bar{q}}-\vec{q}))
\delta(p^0_p+p^0_N - \epsilon_1-\epsilon_2 - q^0)] \label{eq:pn-4}
\end{eqnarray}

We next make a reasonable approximation that
the average energy $\epsilon_1$ ($\epsilon_2$) 
in Eq.(\ref{eq:pn-4})  
is the difference between
the energy of the projectile ($p$) (target $N$) and the removed
parton $q$ ( $\bar{q}$); namely assuming 
\begin{eqnarray}
\delta(p^0_p+p^0_N - \epsilon_1-\epsilon_2 - q^0)
&=&\delta((p^0_p-\epsilon_1)+(p^0_N-\epsilon_2)-q^0) \nonumber \\
 &\sim& \delta(p^0_q+p^0_{\bar{q}}-q^0)\,.
\label{eq:app-2}
\end{eqnarray}
 Then 
 Eq.(\ref{eq:pn-4}) can be written as
\begin{eqnarray}
F^{pN}_{\mu\nu}(p_p,p_N,q)
&=&
\sum_{q}\int d\vec{p}_q d\vec{p}_{\bar{q}}
[F^q_{p_p} (p_q) F^{\bar{q}}_{p_N} (p_{\bar{q}})]
\frac{E_pE_N}{E_qE_{\bar{q}}} \nonumber \\
&&\times 
\{(2\pi)^6 (2E_q) (2E_{\bar{q}})
<\vec{p}_{\bar{q}}\vec{p}_{q}|J_\mu(0)|0><0|J_\nu(0)|\vec{p}_q
\vec{p}_{\bar{q}}> \delta^4(p_q+p_{\bar{q}}-q) \}\,. \nonumber \\
 \label{eq:pn-5}
\end{eqnarray}
The quantity within the bracket $\{...\}$  in the above equation is just
the hadronic tensor $F^{qq}_{\mu\nu}(p_q,p_{\bar{q}})$, defined 
in Eq.(\ref{eq:qq-tensor}),
for the $q\bar{q}$ system.
We thus have
\begin{eqnarray}
F^{pN}_{\mu\nu}(p_p,p_N,q)
&=&\sum_{q}\int d\vec{p}_q d\vec{p}_{\bar{q}} 
[F^q_{p_p} (p_q) F^{\bar{q}}_{p_N} (p_{\bar{q}}) ]
\frac{E_pE_N}{E_qE_{\bar{q}}}F^{qq}_{\mu\nu}(p_q,p_{\bar{q}}) \,.
\label{eq:pN-tensor-1}
\end{eqnarray}
Note that the above simple expression is due to 
 the use of the approximations Eqs.(\ref{eq:delta-f}) and
(\ref{eq:app-2}). The binding effects on the partons in the nucleon and the undetected
fragments $X_p$ and $X_N$ are not treated rigorously.
If we depart from these two simplifications, we then need the spectral function
of the nucleon in terms of parton degrees of freedom to calculate
 DY cross sections. Such information is not available
at the present time.

Substitute Eq.(\ref{eq:pN-tensor-1}) into Eq.(\ref{eq:dy-crst-pn}),
 we then have
\begin{eqnarray}
d\sigma^{pN}
&=&
\sum_{q}\int d\vec{p}_q d\vec{p}_{\bar{q}}
[F^q_{p_p} (p_q) F^{\bar{q}}_{p_N} (p_{\bar{q}})
]\frac{E_pE_N}{E_qE_{\bar{q}}}
\frac{(2\pi)^4}{4[(p_p\cdot p_N)^2-m^2_pm^2_N]^{1/2}} \nonumber \\
&&\times
\{\frac{1}{(2\pi)^6}\frac{d\vec{k}_+}{2E_+}\frac{d\vec{k}_-}{2E_-} 
\frac{1}{q^4}f^{\mu\nu}(k_+,k_-)F^{qq}_{\mu\nu}(p_q,p_{\bar{q}})\}
\label{eq:dy-crst-pn-1}
\end{eqnarray}
The quantity in the bracket $\{ ..\}$ of the above equation is
precisely what is in the bracket $\{ ..\}$ of Eq.(\ref{eq:dy-qbarq-crst}) 
for the $q\bar{q}\rightarrow \mu^+\mu^-$ process. 
Accounting for the difference
in flux factors and extending Eq.(\ref{eq:dy-crst-pn-1}) to
include the
 $q\leftrightarrow \bar{q}$ interchange term,
the full expression of the $p$-$N$ DY process is
\begin{eqnarray}
\frac{d^{pN}(p_p,p_N)}{dq^2}
&=&\sum_{q}\int d\vec{p}_q d\vec{p}_{\bar{q}}
[F^q_{p_p} (p_q) F^{\bar{q}}_{p_N} (p_{\bar{q}})+
F^q_{p_N} (p_q) F^{\bar{q}}_{p_p} (p_{\bar{q}}) ] \nonumber \\
&&\times
\frac{4[(p_q\cdot p_{\bar{q}})^2-m^4_q]^{1/2}}
{4[(p_p\cdot p_N)^2-m_p^2m_N^2)]^{1/2}}
\frac{E_pE_N}{E_qE_{\bar{q}}} 
[\frac{d^{q\bar{q}}(p_q,p_{\bar{q}})}{dq^2}]
\label{eq:dy-pN}
\end{eqnarray}
where $\frac{d^{q\bar{q}}(p_q,p_{\bar{q}})}{dq^2}$ is the
$q$-$\bar{q}$ DY cross section, as defined by  Eq.(\ref{eq:dy-0}).

We now examine the physical meaning of the functions
$F^q_{p_p} (p_q)$ and $F^{\bar{q}}_{p_N} (p_{\bar{q}})$, defined 
in Eqs.(\ref{eq:q-dis})-(\ref{eq:barq-dis}).
The probability of finding a quark $q$ with momentum $\vec{p}_q$ in a
nucleon   state $|\vec{p}_p>$ is defined by 
\begin{eqnarray}
P_{\vec{p}_p}(\vec{p}_q)= \frac{ <\vec{p}_p|b^\dagger_{\vec{p}_q}b_{\vec{p}_q}
|\vec{p}_p|>}
{<\vec{p}_p|\vec{p}_p>}\,.
\label{eq:probab}
\end{eqnarray}
Inserting a complete set of states 
$1=\int d\vec{p}_{X_p}|\vec{p}_{X_p}>
<\vec{p}_{X_p}|$
into the above equation  and using the
definition Eq.(\ref{eq:pn-2}), we then have
\begin{eqnarray}
P_{\vec{p}_p}(\vec{p}_q)&=& \frac{\sum_{X_p}
\int d\vec{p}_{X_p}<\vec{p}_p|b^\dagger_{\vec{p}_q}|\vec{p}_{X_p}>
<\vec{p}_{X_p}|b_{\vec{p}_q}|\vec{p}_p|>}
{<\vec{p}_p|\vec{p}_p>} \nonumber \\
&=&\frac{\sum_{X_p}
\int d\vec{p}_q|\phi_{\vec{p}_p}(\vec{p}_{\bar{q}} ,\vec{p}_{X_p})|^2
\delta(\vec{p}_p-\vec{p}_q -
\vec{p}_{X_p})\delta(\vec{p}_p-\vec{p}_q -\vec{p}_{X_p})}
{<\vec{p}_p|\vec{p}_p>} \,.\nonumber 
\end{eqnarray}
With our normalization $<\vec{p}_p|\vec{p}_p>= \delta(\vec{p}_p-\vec{p}_p)$, 
the above equation becomes
\begin{eqnarray}
P_{\vec{p}_p}(\vec{p}_q)
&=&[\frac{\sum_{X_p}
\int d\vec{p}_q|\phi_{\vec{p}_p}(\vec{p}_{\bar{q}} ,\vec{p}_{X_p})|^2
\delta(\vec{p}_p-\vec{p}_q -
\vec{p}_{X_p})] \delta(\vec{p}_p-\vec{p}_p)}{\delta(\vec{p}_p-\vec{p}_p)}
\nonumber \\
&=& \sum_{X_p}
\int d\vec{p}_q|\phi_{\vec{p}_p}(\vec{p}_{\bar{q}} ,\vec{p}_{X_p})|^2
\delta(\vec{p}_p-\vec{p}_q -
\vec{p}_{X_p})
\end{eqnarray}
Clearly  $F^{q}_{\vec{p}_p}(\vec{p}_q)$ of Eq.(\ref{eq:q-dis}) is identical
to $P_{\vec{p}_p}(\vec{p}_q)$ given above and is the probability of finding
a quark with momentum $\vec{p}_q$ in a  nucleon  moving with a momentum
$\vec{p}_p$. Thus $F^{q}_{\vec{p}_p}(\vec{p}_q)$ is the parton  
distribution function in the nucleon.

\begin{figure}[t]
\begin{center}
\includegraphics[clip,width=0.5\textwidth]{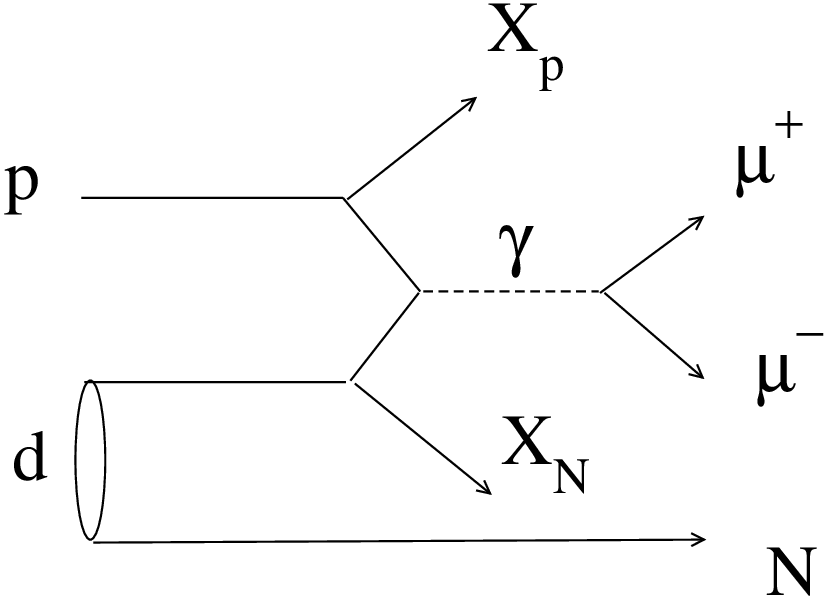}
\caption{Impulse approximation of $p+d$
Drell-Yan process.}
 \label{fig:dy-mech-imp}
\end{center}
\end{figure}

\section{$p$-$d$ DY cross section}

For  the
$p(p_p) + d(p_d) \rightarrow \mu^+(k_+) +
\mu^-(k_-)$ 
DY process, Eq.(\ref{eq:dy-crst}) gives
\begin{eqnarray}
d\sigma^{pd} =
\frac{(2\pi)^4}{4[(p_p\cdot p_d)^2
-m^2_pm^2_d]^{1/2}}
\{\frac{1}{(2\pi)^6}\frac{d\vec{k}_+}{2E_+}\frac{d\vec{k}_-}{2E_-}
\frac{1}{q^4}f^{\mu\nu}(k_+,k_-)F^{pd}_{\mu\nu}(p_p,p_d,q)\} \,.
\label{eq:dy-crst-pd}
\end{eqnarray}
We assume that the $p$-$d$ DY
process occurs only on each of the nucleons in the deuteron, as
illustrated in Fig.\ref{fig:dy-mech-imp}.
In this impulse approximation, the hadronic tensor for a
deuteron target
can be obtained by simply
extending 
 Eq.(\ref{eq:pN-tensor}) for the $p$+$N$ to include a
 spectator nucleon state $|p_s>$
in the sum over the final hadronic states. We thus have
\begin{eqnarray}
F^{pd}_{\mu\nu}(p_p,p_d,q)
&=&
(2\pi)^6 (2E_p)(2E_d) \sum_{X_p,X_N} \int d\vec{p}_s
 d\vec{p}_{X_p}  d\vec{p}_{X_N}
\delta^4(p_p+p_d-p_{X_p}-p_{X_N}-p_s-q)
\nonumber \\
&&\times [<\Phi_{p_d}p_p |J_\mu(0)|p_{X_p} p_{X_N} p_s>
<p_s p_{X_N} p_{X_p}|J_\nu(0)|p_p\Phi_{p_d}>] \,,
\label{eq:pd-tensor-0}
\end{eqnarray}
where $|\Phi_{p_d}>$ is the state of a deuteron moving with a 
momentum $p_d$. Within  the framework of
relativistic quantum mechanics\cite{kp-book}, we can expand
the deuteron state in terms of two-nucleon plane-wave state
 $|\vec{p}_N\vec{p}_2>$
\begin{eqnarray}
|\Phi_{p_d}> = \int d\vec{p}_N d\vec{p}_2 \Phi_{p_d}(\vec{p}_N)
\delta(\vec{p}_d-\vec{p}_N-\vec{p}_2)
|\vec{p}_N\vec{p}_2>\,.
\label{eq:d-wf}
\end{eqnarray}
Keeping only the contributions which are due to a
parton in $|\vec{p}_N>$ of the above expansion 
and a parton from projectile state $|\vec{p}_p>$, we have
\begin{eqnarray}
&&<\Phi_{p_d}p_p |J_\mu(0)|p_{X_p} p_{X_N} p_s> \nonumber \\
&& =\int  d\vec{p}_2 \int d \vec{p}_N \Phi^*(\vec{p}_N)
\delta(\vec{p}_d-\vec{p}_N-\vec{p}_2) <p p_N | J_\mu(0)|p_{X_p}p_{X_N}>
<\vec{p}_2|\vec{p}_s> \nonumber \\
&& = \int d \vec{p}_N \Phi^*(\vec{p}_N)
\delta(\vec{p}_d-\vec{p}_N-\vec{p}_s) <p p_N | J_\mu(0)|p_{X_p}p_{X_N}> \,.
\label{eq:c-mxd}
\end{eqnarray}
By using Eq.(\ref{eq:c-mxd}), Eq.(\ref{eq:pd-tensor-0})
can then be written as
\begin{eqnarray}
F^{pd}_{\mu\nu}(p_p,p_d,q)
&=&\int d\vec{p}_N |\Phi_{p_d}(p_N)|^2\frac{(2E_p)(2E_d)}{(2E_p)(2E_N)}
\nonumber \\
&&\times \{
(2\pi)^6 (2E_p)(2E_N)
\times \sum_{X_p,X_N}\int d\vec{p}_{X_p} d\vec{p}_{X_N}
\delta^4(p_p+p_N-p_{X_p}-p_{X_N}-q) \nonumber \\
&&\times
[<p_Np_p |J_\mu(0)|p_{X_p} p_{X_N}><p_{X_N} p_{X_p}|J_\nu(0)|p_pp_N>]\} \,.
\label{eq:pd-tensor}
\end{eqnarray}
We see that the quantity within the  bracket $\{..\}$ in the
above equation is identical to
$F^{pN}_{\mu\nu}(p_p,p_N,q)$ of Eq.(\ref{eq:pN-tensor}).
We then have 
\begin{eqnarray}
F^{pd}_{\mu\nu}(p_p,p_d,q)
&=&
\int d\vec{p}_N \rho_{p_d}(\vec{p}_N)
\frac{E_p E_d}{E_pE_N} F^{pN}_{\mu\nu}(p_p,p_N,q) \,, \label{eq:pd-t11}
\end{eqnarray}
 where
\begin{eqnarray}
 \rho_{p_d}(\vec{p}_N) = |\Phi_{p_d}(p_N)|^2.
\end{eqnarray}
By using Eq.(\ref{eq:d-wf}), one can show that $\rho_{p_d}(\vec{p}_N)$
is the nucleon momentum distribution in a deuteron moving
with a momentum $p_d$ :
\begin{eqnarray}
\rho_{p_d}(\vec{p}_N) 
=\frac{<\Phi_{p_d}|b^\dagger_{\vec{p}_N}b_{\vec{p}_N}|\Phi_{p_d}>}
{<\Phi_{p_d}|\Phi_{p_d}>} \,.
\label{eq:rho-n}
\end{eqnarray}
We will present formula for calculating $\rho_{p_d}(\vec{p}_N)$ in section V.

By using Eq.(\ref{eq:pd-t11}), Eq.(\ref{eq:dy-crst-pd}) becomes
\begin{eqnarray}
d\sigma^{pd} &=&\frac{(2\pi)^4}{4[(p\cdot p_d)^2
-m^2_pm^2_d]^{1/2}}\sum_{N=p,n}\int  d\vec{p}_N \rho_{p_d}(\vec{p}_N) 
\frac{E_p E_d}{E_pE_N}
\nonumber \\
&&\times\{\frac{1}{(2\pi)^6}\frac{d\vec{k}_+}{2E_+}\frac{d\vec{k}_-}{2E_-}
\frac{1}{q^4}f^{\mu\nu}(k_+,k_-)
 F^{pN}_{\mu\nu}(p_p,p_N,q) \} \,.
\label{eq:dy-crst-pd-1}
\end{eqnarray}
The quantity within the bracket $\{..\}$  of the above equation is
exactly what is in the bracket $\{..\}$ of
 Eq.(\ref{eq:dy-crst-pn}). Accounting for the difference in flux factor,
we obviously can write
\begin{eqnarray}
\frac{d^{pd}(p,p_d)}{dq^2}
&=&\sum_{N=p,n}\int  d\vec{p}_N \rho_{p_d}(\vec{p}_N)
\frac{[(p\cdot p_N)^2-m^2_pm^2_N]^{1/2}}
{[(p\cdot p_d)^2-m_p^2m_d^2)]^{1/2}}
\frac{E_p E_d}{E_pE_N} \frac{d^{pN}(p,p_N)}{dq^2} \,,
\label{eq:dy-pd-000}
\end{eqnarray}
where $ \frac{d^{pN}(p,p_N)}{dq^2}$ is given in Eq.(\ref{eq:dy-pN}).
Substituting Eq.(\ref{eq:dy-pN}) into Eq.(\ref{eq:dy-pd-000}),
we have
\begin{eqnarray}
\frac{d^{pd}(p,p_d)}{dq^2}
&=&\sum_{N=p,n}\int  d\vec{p}_N \rho_{p_d}(\vec{p}_N) 
\{\sum_{q}\int d\vec{p}_q d\vec{p}_{\bar{q}}
\frac{[(p_q\cdot p_{\bar{q}})^2-m^4_q]^{1/2}}
{[(p\cdot p_d)^2-m_p^2m_d^2)]^{1/2}}
\frac{E_p E_d}{E_qE_{\bar{q}}} \nonumber \\
&&\times
[F^q_{p_p} (p_q) F^{\bar{q}}_{p_N} (p_{\bar{q}})+
F^q_{p_N} (p_q) F^{\bar{q}}_{p_p} (p_{\bar{q}}) ] 
[\frac{d^{q\bar{q}}(p_q,p_{\bar{q}})}{dq^2}]\}\,.
\label{eq:dy-pd}
\end{eqnarray}
Clearly the input to our calculations of the $p$-$d$ DY cross sections
are the nucleon momentum
distribution $\rho_{p_d}(\vec{p}_N)$, the parton distribution
functions $F^q_{\vec{p}}(\vec{p}_q)$ and $F^{\bar{q}}_{\vec{p}}(\vec{p}_q)$,
and the elementary partonic $q\bar{q}\rightarrow \mu^+\mu^-$
cross section $\frac{d^{q\bar{q}}(p_q,p_{\bar{q}})}{dq^2}$ 
defined by  Eq.(\ref{eq:dy-0}).

Here we note that within the impulse approximation, as illustrated in
Fig.\ref{fig:dy-mech-imp}, the binding on the nucleon in the deuteron is
treated accurately if $\rho_{p_d}(\vec{p}_N)$ is calculated from realistic nucleon-nucleon
potentials which give high precision fits to the NN scattering data and the  deuteron
properties. In the considered framework of Relativistic Quantum Mechanics, the
nucleon emitting a parton is on its mass-shell, while the interacting two-nucleons system 
is off the energy-shell of the deuteron:
 $\sqrt{\vec{p}^{\,\,2}_1+m^2_N}+\sqrt{\vec{p}^{\,\,2}_2+m^2_N}\neq \sqrt{\vec{p}^{\,\,2}_d + m^2_d}$. 
Precisely because of this choice of formulation, the parton
distribution in Eq.(\ref{eq:dy-pd}) is the same as that in the free nucleon.
If one chooses a theoretical framework within which
 the nucleon in the deuteron can be off its
mass-shell $p^2\neq m^2_N$, then this simple relation is lost and one needs
addtional theoretical assumptions to  extract
the $\bar{d}/\bar{u}$ ratios in the proton from $p$-$p$ and $p$-$d$ DY data.

\section{Numerical Procedures}

In this section, we develop numerical procedures for applying the
formula presented in previous sections to investigate the
Fermi motion effect on the ratio
 $R_{pd/pp}=\sigma^{pd}/(2\sigma^{pp})$ between the $p$-$d$ and
$p$-$p$ DY cross sections. Our first task is 
to relate our momentum variables $p_p$, $p_T$ and $q$
to the variables used in the analysis\cite{peng2001} of the available data.
This will be given in the
first subsection V.A. The procedures for calculating DY cross sections 
are given for $p$-$p$ in
subsection V.B and $p$-$d$ in subsection V.C. The calculations of
momentum distributions $\rho_{p_d}(\vec{p}_N)$ of
a moving deuteron  are described in subsection V.D.
 
\subsection{Kinematic variables for DY cross sections}
It is common\cite{peng2001} to use the co-linear approximation 
to define the parton momentum:
\begin{eqnarray}
p_{q_p}= x_1 p_p \label{eq:x1x2-a} \,,\\
p_{q_T}=x_2 p_T  \label{eq:x1x2-b} \,, 
\end{eqnarray}
where $p_{q_p}$ ($ p_{q_T})$ is the momentum of a parton in the 
projectile (target), 
and $x_1$ and $x_2$ are scalar numbers.
The momentum $q$ of the virtual photon in the $q\bar{q} \rightarrow \gamma
\rightarrow \mu^+ \mu^-$ can then be written as
\begin{eqnarray}
q&=&p_q+p_{\bar{q}} \nonumber \\
&=& x_1p_p+x_2p_T \,. \label{eq:q}
\end{eqnarray}
In the considered very  high energy region $E_p > 100 $ GeV,
the  masses of projectile ($p^2_p=m^2_p$) and target ($p^2_T=m^2_T$)
can be neglected and we have
\begin{eqnarray} 
&&s=(p_p+p_T)^2 \sim  2p_p\cdot p_T \,,\nonumber \\ 
&&p_T\cdot q \sim x_1 p_p\cdot p_T \,, \nonumber \\ 
&& p_p\cdot q \sim x_2 p_p\cdot p_T \,. \nonumber
\end{eqnarray}
The above relations give
\begin{eqnarray}
x_1&\sim&\frac{2q\cdot p_T}{s}  \label{eq:x1-0}\,,\\
x_2&\sim&\frac{2q\cdot p_p}{s}  \label{eq:x2-0}\,.
\end{eqnarray}

It is most convenient to perform calculations in terms of $x_1$ and $x_2$
in the center of mass (c.m.)
 system in which the projectile is in the  $z$ direction
and the target in
-$z$ direction:
\begin{eqnarray}
p_p &=& (\sqrt{p^2+m^2_p},0,0,p)\sim (p,0,0,p) \label{eq:pp-cm}\,, \\
p_T &=& (\sqrt{p^2+m^2_T},0,0,-p) \sim (p,0,0,-p) \label{eq:pt-cm}\,.
\end{eqnarray}
With the choices  Eqs.(\ref{eq:pp-cm})-(\ref{eq:pt-cm}),
we 
\begin{eqnarray}
s&=& (p_p+p_T)^2 \sim 4p^2 \,, \nonumber \\
M^2 &=& q^2= (x_1p_p+x_2p_T)^2 \sim 4 x_1x_2 p^2 \,. \nonumber
\end{eqnarray}
The above two equations leads the simple relation
\begin{eqnarray}
x_1x_2\sim \frac{M^2}{s} \label{eq:x1x2-ms} \,.
\label{eq:mx1x2}
\end{eqnarray}
By using Eqs.(\ref{eq:x1-0})-(\ref{eq:x2-0}) and
Eqs.(\ref{eq:pp-cm})-(\ref{eq:pt-cm}), 
we can define a useful variable $x_F$ 
\begin{eqnarray}
x_F&=&x_1-x_2
\sim\frac{2(p_T-p_p)\cdot q}{s} \nonumber \\
&\sim&\frac{2\sqrt{s}\hat{z} \cdot \vec{q}}{s}
=\frac{2\sqrt{s}\hat{p}_p \cdot \vec{q}}{s}\,.
\end{eqnarray}
In the notation of Ref.\cite{peng2001}, we write
\begin{eqnarray}
x_F\sim\frac{p^\gamma_{\parallel}}{\sqrt{s}/2}\,,
\label{eq:xf}
\end{eqnarray}
where $p^\gamma_{\parallel}=\hat{p}_p \cdot \vec{q}$ is
clearly  the longitudinal momentum of the
intermediate photon with respect to the projectile in the
c.m. frame.
Experimentally, $s$, $M$, $x_F$, and $d\sigma/(dM dx_F)$ are
 measured\cite{peng-cm}.
With the relation Eq.(\ref{eq:x1x2-ms}), we certainly can determine
the corresponding $x_1$, $x_2$ and $d\sigma/(dx_1dx_2)$.
We thus will only  give the expression of $d\sigma/(dx_1dx_2)$ in the
following subsections.

Before we proceed further, 
it is necessary to discuss how our formula can be expressed in terms of
$x_1$ and $x_2$.
As given in Eqs.(\ref{eq:x1x2-a})-(\ref{eq:x1x2-b}),
 these variables define the fraction of
the hadron momentum carried by the emitted parton.
 For the projectile proton, this is unambiguous 
$x_1=p_q/p_p = p^z_q/p^z_p$ since
we choose $\vec{p}$ in z-direction. For the
deuteron target, $x_2$ could be identified 
with the deuteron or the nucleons inside the deuteron since
in the above derivations hadron masses are all neglected.
Since the impulse approximation is used in our derivations,
we identify $x_2$ with the mean nucleon momentum $<p_N>$ in the deuteron.
As will be seen from the realistic nucleon momentum  distribution to be
presented in  Fig.\ref{fig:rho-v14}, $ <p_N> \sim |\vec{p_d}|/2$ as 
expected naively.

\subsection{Calculation of $p$-$p$ DY cross sections $d\sigma^{pp}/dx_1dx_2$}
We now note that with the variables $x_1$ and $x_2$ defined above
the flux factor associated with Eq.(\ref{eq:dy-pN}) become 1.
Substituting Eq.(\ref{eq:dy-0}) into Eq.(\ref{eq:dy-pN}), the
DY cross section for 
 $p(p_p)+p(p_N) \rightarrow \mu^+\mu^-$ is then calculated from
\begin{eqnarray}
\frac{d^{pN}(p_p,p_N)}{dq^2}
&=&\sum_{q}\int d\vec{p}_q d\vec{p}_{\bar{q}}
[F^q_{p_p} (p_q) F^{\bar{q}}_{p_N} (p_{\bar{q}})+
F^q_{p_N} (p_q) F^{\bar{q}}_{p_p} (p_{\bar{q}}) ] \nonumber \\
&&\times
\frac{4\pi\alpha^2}{9q^2}e^2_q\delta(q^2-(p_q+p_{\bar{q}}))
\label{eq:dy-pN-c0} \,.
\end{eqnarray}
In the chosen center of mass frame, defined by 
Eqs.(\ref{eq:pp-cm})-(\ref{eq:pt-cm}),
 let us consider $\bar{q}$ in the target proton moving with
a momentum 
$p_N=(p^z_N,0,0,p^z_N)$.
In the 
precise co-linear approximation, only the z-component of the $\bar{q}$
momentum is defined by $p^z_N$. We thus  write
$\vec{p}_{\bar{q}}=(\vec{p}_{\bar{q},T}, p^z_{\bar{q}})$ where
\begin{eqnarray}
p^z_{\bar{q}} = x_2 p^z_N \,, \label{eq:pz-x2}
\end{eqnarray}
and  the transverse component $\vec{p}_{\bar{q},T}$ can be arbitrary.
The integration over the $\bar{q}$ momentum distribution in the
target $N$ can then be written as
\begin{eqnarray}
 \int d\vec{p}_{\bar{q}}F^{\bar{q}}_{p_N}(\vec{p}_{\bar{q}})
&=&\int dp^z_{\bar{q}} \int d\vec{p}_{\bar{q},T} F^{q}_{p}(p^z_{\bar{q}},\vec{p}_{\bar{q},T}) \nonumber \\
&=& p^z_N\int  dx_2\int d\vec{p}_{\bar{q},T}
F^{q}_{p}(x_2p^z_N,\vec{p}_{\bar{q},T}) \nonumber \\
&=& \int dx_2 f^{\bar{q}}_{p_N}(x_2) \,,
\end{eqnarray}
with
\begin{eqnarray}
f^{\bar{q}}_{p_N}(x_2 ) = p^z_N \int d\vec{p}_{\bar{q},T} 
F^{q}_{p_N}(x_2 p^z_N,\vec{p}_{\bar{q},T}) \,.
\label{eq:pdf-1}
\end{eqnarray}
We identify 
$f^{\bar{q}}_{p_N}(x_2 )$ with the parton distribution functions (PDF)
determined by several groups\cite{lai97,mart96,gluc95,plot95,ceteq5m}.
To compare with the results of Ref.\cite{peng2001},
we use the  PDF of CETEQ5m\cite{ceteq5m}.

Similarly, we can define for the projectile proton
\begin{eqnarray}
 \int d\vec{p}_{q}F^{q}_{p_p}(\vec{p}_{q})
= \int dx_1 f^q_{p_p}(x_1)\,,
\label{eq:pdf-2}
\end{eqnarray}
where $f^q_{p_p}(x_1)$ is defined by the same Eq.(\ref{eq:pdf-1}) with
$q$ replaced by $\bar{q}$ and $x_2$ by $x_1$.
By using Eqs.(\ref{eq:pdf-1}) and (\ref{eq:pdf-2}), 
Eq.(\ref{eq:dy-pN-c0}) can be written as
\begin{eqnarray}
\frac{d^{pN}(p_p,p_N)}{dq^2}
&=&\sum_{q}\int dx_1 dx_2
[f^q_{p_p} (x_1) f^{\bar{q}}_{p_N} (x_2)+
f^q_{p_N} (x_1) f^{\bar{q}}_{p_p} (x_2) ] \nonumber \\
&&\times
\frac{4\pi\alpha^2}
{9q^2}e^2_q\delta(q^2-(p_q+p_{\bar{q}})^2) \,.
\label{eq:dy-pN-c}
\end{eqnarray}
By integrating over $dq^2$, the above equation leads to
\begin{eqnarray}
\frac{d^{pN}(p_p,p_N)}{dx_1dx_2}&=&
\sum_{q}\frac{4\pi\alpha^2}{9(p_q+p_{\bar{q}})^2}e^2_q[f^q_{p_p} (x_1) f^{\bar{q}}_{p_N} (x_2)+
f^q_{p_N} (x_1) f^{\bar{q}}_{p_p} (x_2) ] \,. 
\label{eq:dy-pp-c}
\end{eqnarray}
which is the same as 
Eq.(\ref{eq:dy-exp}) used in the analysis of Ref.\cite{peng2001}
since $(p_q+p_{\bar{q}})^2= q^2=M^2$ for the partonic
process $q\bar{q}\rightarrow \gamma$.

We only consider up ($u$) and down ($d$) 
quarks in the proton.
Eq.(\ref{eq:dy-pp-c}) for the $p$-$p$ DY cross section
 then obviously takes the following
form
\begin{eqnarray}
\frac{d^{pp}(p,p_p)}{dx_1dx_2}&=&\frac{4\pi\alpha^2}{9M^2}
[\frac{4}{9}(f^{\bar{u}}_p(x_1)f^u_{p_p}(x_2)+f^{{u}}_p(x_1)
f^{\bar{u}}_{p_p}(x_2)) \nonumber \\
&& +\frac{1}{9}(f^{\bar{d}}_p(x_1)f^{{d}}_{p_p}(x_2)+
f^{d}_p(x_1)f^{\bar{d}}_{p_p}(x_2))] \,.
\label{eq:dy-pp-cc}
\end{eqnarray}

\subsection{Calculation of $pd$ DY cross sections of $d\sigma^{pd}/dx_1dx_2$}
From Eq.(\ref{eq:dy-pd}), we see that the calculations of
 $p$-$d$ DY cross sections involve the convolution of
the  parton distribution associated
with a nucleon with momentum $\vec{p}_N$ over
 the nucleon momentum distribution $\rho_{p_d}(\vec{p}_N)$ in
a moving deuteron.
To see this more clearly, we consider 
the contribution from an anti-quark in a nucleon $N$ of the  deuteron 
target
 and a quark in the projectile proton. 
By using Eq.(\ref{eq:dy-0}), this particular contribution to
Eq.(\ref{eq:dy-pd})
is
\begin{eqnarray}
\frac{d^{pd}(p,p_d)}{dq^2}
&=&
\int d\vec{p}_q\int d\vec{p}_{\bar{q}}
\{\frac{[(p_q\cdot p_{\bar{q}})^2-m^4_q]^{1/2}}
{[(p\cdot p_d)^2-m_p^2m_d^2)]^{1/2}}
\frac{E_N(p) E_d(p_d)}{E_q(p_q)E_q(p_{\bar{q}})}\} \nonumber \\
&& \times \int  d\vec{p}_N \rho_{p_d}(\vec{p}_N)
[f^{q}_{p}(\vec{p}_q)f^{\bar{q}}_{p_N}(\vec{p}_{\bar{q}}]
\frac{4\pi\alpha^2}{9q^2}e^2_q\delta(q^2-(p_q+p_{\bar{q}}))\,.
\label{eq:pd-dy-000}
\end{eqnarray}
Bu using the variables $x_1$ and $x_2$ defined in subsection V.A,
 the ratio between flux factors in 
the $\{..\}$ bracket of the above equation  is
close to 1. We thus only need to consider
\begin{eqnarray}
\frac{d^{pd}(p,p_d)}{dq^2}
&=&
\int d\vec{p}_q \int d\vec{p}_{\bar{q}}
\int  d\vec{p}_N \rho_{p_d}(\vec{p}_N)
[f^{q}_{p}(\vec{p}_q)f^{\bar{q}}_{p_N}(\vec{p}_{\bar{q}}] \nonumber \\
&&\times \frac{4\pi\alpha^2}{9q^2}e^2_q\delta(q^2-(p_q+p_{\bar{q}})^2)
\label{eq:pd-dy-01}
\end{eqnarray}

We now define the momentum fraction $x^N_2$ of
a parton associated with a nucleon with momentum $p_N$ in the deuteron
\begin{eqnarray}
p_{\bar{q}}=x^N_2 p^z_N\,. \label{eq:x2n}
\end{eqnarray}
By using the definitions Eq.(\ref{eq:x2n}) for $x^N_2$ and Eqs.(\ref{eq:pdf-1})-(\ref{eq:pdf-2}) for 
parton distributions, 
Eq.(\ref{eq:pd-dy-01}) can be written as
\begin{eqnarray}
\frac{d^{pd}(p,p_d)}{dq^2}
&=&\int dx_1\int dx^N_2\int d\vec{p}_{N}                                
 \rho_{p_d}(\vec{p}_{N})
[f^{q}_{p}(x_1)f^{\bar{q}}_{p_N}(x^N_2)] \nonumber \\
&&\times \frac{4\pi\alpha^2}{q^2}e^2_q\frac{1}{9}
\delta(q^2-(p_q+p_{\bar{q}})^2)
\label{eq:dy-pd-00}
\end{eqnarray}

Similar to the $pp$ case, $\vec{p}_d =(p^z_d, \vec{p}_{d,T}=0)$ 
is chosen to be on the z-direction of the c.m.frame.
As discussed above,  we need to find a way to relate $x^N_2$ to $x_2$ which
is determined by experimental variables $M$, $s$ and $x_f$ through
the relations: $x_1x_2=M^2/s$ and
$x_f=x_1-x_2 = p^\gamma_\parallel/(\sqrt{s}/2)$.
In the impulse approximation we have used to derive the $p$-$d$ DY cross section,
it is assumed that the
parton  is emitted from one of the  nucleons in the deuteron, as illustrated in 
Fig.\ref{fig:dy-mech-imp}. Thus  it is
reasonable to choose $p_{\bar{q}}=x_2 p^z_{ave}$ where
$p^z_{ave}$ is the averaged nucleon momentum in deuteron. Combining this assumption and
Eq.(\ref{eq:x2n}), we then have the following relation
\begin{eqnarray}
x^N_2 =x_2\frac{p^z_{ave}}{p^z_N}\,. \label{eq:x2n-a}
\end{eqnarray} 
In section VI, we will determine  $p^z_{ave}$ from investigating 
$\rho_{p_d}(\vec{p}_N)$ generated from realistic deuteron wavefunctions.

By using Eq.(\ref{eq:x2n-a}) to change the integration variable, we can write
Eq.(\ref{eq:dy-pd-00}) as
\begin{eqnarray}
\frac{d^{pd}(p,p_d)}{dq^2}
&=&\int dx_1\int dx_2\int d\vec{p}_{N}
 \rho_{p_d}(\vec{p}_{N})\frac{p^z_{ave}}{p^z_N}
[f^{q}_{p}(x_1)f^{\bar{q}}_{p_N}(\frac{x_2p^z_{ave}}{p^z_N})] \nonumber \\
&&\times \frac{4\pi\alpha^2}{q^2}e^2_q\frac{1}{9}\delta(q^2-(p_q+p_{\bar{q}})^2)
\end{eqnarray}
Integrating over $q^2$ on both sides of the above equation, we obtain
\begin{eqnarray}
\frac{d^{pd}(p,p_d)}{dx_1dx_2}
&=& [f^{q}_{p}(x_1) \Gamma^{\bar{q}}_{p_d,N}(x_2)]
\frac{4\pi\alpha^2}{9q^2}e^2_q \,,
\label{eq:dy-pd-x1x2}
\end{eqnarray}
where the $\bar{q}$ contribution is isolated in
\begin{eqnarray}
\Gamma^{\bar{q}}_{p_d,N}(x_2)=
\int d\vec{p}_{N}
 \rho_{p_d}(\vec{p}_{N})[\frac{p^z_{ave}}{p^z_N}
 f^{\bar{q}}_{p_N}(\frac{x_2p^z_{ave}}{p^z_N})] \,. 
\label{eq:pdf-ave}
\end{eqnarray}
 
The derivation of Eq.(\ref{eq:pdf-ave}) can be extended to have $q$ in deuteron and $\bar{q}$
in the projectile proton.
 We finally obtain
\begin{eqnarray}
\frac{d^{pd}(p,p_d)}{dx_1dx_2}
&=&\sum_{q}\{ 
f^{q}_{p}(x_1)
[\sum_{N=p,n}\Gamma^{\bar{q}}_{p_d,N}(x_2)]+ 
f^{\bar{q}}_{p}(x_1)[\sum_{N=p,n} \Gamma^{q}_{p_d,N}(x_2)]\}
\frac{4\pi\alpha^2}{9q^2}e^2_q \,.
\label{eq:dy-pd-c}
\end{eqnarray}

We use the charge symmetry to calculate the parton distribution functions
for the neutron from that of proton:
 $f^d_n =f^u_p$,
 $f^u_n =f^d_p$, 
$f^{\bar{d}}_n =f^{\bar{u}}_p$, 
$f^{\bar{u}}_n =f^{\bar{d}}_d$.
Furthermore $\rho_{p_d}(p_N)$ is the same for neutron and proton.
Including the charges for $u$ and $d$ quarks appropriately,
Eq.(\ref{eq:dy-pd-c}) can be written as
\begin{eqnarray}
\frac{d^{pd}(p,p_d)}{dx_1dx_2}
&=&\frac{4\pi\alpha^2}{9q^2}[\frac{4}{9}f^u_p(x_1)+\frac{1}{9}f^d_p(x_1)][\Gamma^{\bar{u}}_{p_d,p}(x_2)
+\Gamma^{\bar{d}}_{p_d,p}(x_2)] \nonumber \\
&& +[\frac{4}{9}f^{\bar{u}}_p(x_1)+
\frac{1}{9}f^{\bar{d}}_p(x_1)][\Gamma^{u}_{p_d,p}(x_2)
+\Gamma^{d}_{p_d,p}(x_2)] \,.
\label{eq:dy-pd-cc}
\end{eqnarray}

\subsection{Calculation of $\rho_{p_d}(\vec{p}_N)$}
To calculate the $p$-$d$ DY cross section Eq.(\ref{eq:dy-pd}), we need to 
generate the nucleon momentum
distribution $\rho_{P_d}(\vec{p}_N)$ defined by
Eq.(\ref{eq:rho-n}).
Since the available realistic nucleon-nucleon potentials only
provide the wavefunctions in the rest frame, we boost
these wavefunctions  within the relativistic quantum mechanics
proposed by Dirac, as reviewed in Ref.\cite{kp-book}.
We note here that the parton distributions defined by
Eqs.(\ref{eq:q-dis})-(\ref{eq:barq-dis}) or
Eqs.(\ref{eq:pdf-1})-(\ref{eq:pdf-2})
are also 
within the same framework. We thus have a consistent 
relativistic description of both the nucleonic 
structure of the deuteron and the partonic structure
of the nucleon. Such a consistency is essential
in investigating the Fermi motion effect on DIS and DY processes.

Starting with a mass operator (Hamiltonian in
the rest frame) which defines the 
mass and the wavefunction $\chi_d(\vec{k})$ of the
deuteron in its rest frame, one can generate the wavefunction $\Phi_{p_d}$ of
a moving deuteron with momentum $p_d$ using one of the three 
possible forms\cite{dirac,kp-book}  of the relativistic quantum mechanics. 
Furthermore, one can identify
 $\chi_d(\vec{k})$ with the deuteron wavefunctions generated from the available
realistic nucleon-nucleon potentials, as discussed in Refs.\cite{coester-a,polyzou-a}.
Here we consider the Instant-Form (IF) and Light-Form (LF).
The formula for calculating $\rho_{p_d}(\vec{p}_N)$
can be derived by using the procedures presented in
Ref.\cite{kp-book} and will not be  detailed here. We will simply write
down 
the formula used in our calculations.

\subsubsection{Instant Form}
The deuteron wavefunction $\chi_d(\vec{k})$ in the $\vec{p}_d=0$ rest frame
 is normalized as
\begin{eqnarray}
\int d\vec{k}|\chi_d(\vec{k})|^2=1\,.\nonumber
\end{eqnarray}
If we neglect the interaction in the Lorentz Boost and follow the procedures
detailed  in Ref.\cite{kp-book}, it is straightforward to show that 
Eqs.(\ref{eq:d-wf}) and (\ref{eq:rho-n}) lead to
\begin{eqnarray}
\rho_{p_d}(\vec{p}) &=&
\frac{<\Phi_{p_d}|b^\dagger_{\vec{p}}b_{\vec{p}}|\Phi_{p_d}>}
{<\Phi_{p_d}|\Phi_{p_d}>} \nonumber \\
 &=& |\frac{\partial(\vec{p}_d,\vec{k})}
{ \partial(\vec{p},\vec{p}_2)}| |\chi_d(\vec{k})|^2 \,,
\label{eq:fp-if}
\end{eqnarray}
with
\begin{eqnarray}
|\frac{\partial(\vec{p}_d,\vec{k})}{ \partial(\vec{p},\vec{p}_2)}|
&=&\frac{
\omega_m(\vec{k})\omega_m(\vec{k})\omega_{M_0}(\vec{p}_d)}
{ \omega_m(\vec{p})
\omega_m(\vec{p}_2)M_0} \label{jacob-if} \,, \\
\end{eqnarray}
where $m$ is the nucleon mass, $\omega_m(\vec{k}) = \sqrt{m^2+\vec{k}^{\,\,2}}$, and
\begin{eqnarray} 
\vec{p}_2&=&\vec{p}_d-\vec{p} \label{eq:if-k1}\,, \\
\vec{k}&=&\vec{p}+\frac{\vec{p}_d}{M_0}[\frac{\vec{p}_d\cdot\vec{p}}{M_0+H_0}
-\omega_m(\vec{p})] \label{eq:if-k2} \,, \\
H_0&=&\omega_m(\vec{p})+\omega_m(\vec{p}_2) \label{eq:if-k3} \,, \\
M_0&=&[H^2_0-\vec{p}_d^{\,2}]^{1/2} \label{eq:if-k4} \,.
\end{eqnarray}

\subsubsection{Light Form}
The Light-Form (LF) momentum is defined by
\begin{eqnarray}
 \tilde{p}=(p^+,\vec{p}_\perp)\,, \nonumber
\end{eqnarray}
where
\begin{eqnarray}
p^+&=& \omega_m(p) +p^3\,, \nonumber \\
\vec{p}_\perp&=&(p^1,p^2)\,. \nonumber
\end{eqnarray}
By using the above definitions for all momenta in Eqs.(\ref{eq:d-wf}) and (\ref{eq:rho-n})
and following the procedures
detailed  in Ref.\cite{kp-book},
one can show that the nucleon momentum distribution in LF is
\begin{eqnarray}
\rho^{LF}_{\tilde{p}_d}(\tilde{p})
 &=&
\frac{<\Phi_{\tilde{p}_d}|b^\dagger_{\tilde{p}}b_{\tilde{p}}|\Phi_{\tilde{p}_d}>}
{<\Phi_{\tilde{p}_d}|\Phi_{\tilde{p}_d}>} \nonumber \\
&=& 
|\frac{\partial(\tilde{p}_d,\vec{k})}
{ \partial(\tilde{p},\tilde{p}_2)}| |\chi_d(\vec{k})|^2 \,,
\label{q:fp-lf}
\end{eqnarray}
where
\begin{eqnarray}
|\frac{\partial(\tilde{p}_d,\vec{k})}{ \partial(\tilde{p},\tilde{p}_2)}|
&=&\frac{
\omega_m(\vec{k})\omega_m(\vec{k})p^+_d}
{p^+p^+_2 M_0} \label{eq:jaco-lf}
\end{eqnarray}
with
\begin{eqnarray}
\tilde{p}_2&=&\tilde{p}_d-\tilde{p} \label{eq:lf-k1} \,,  \\
\vec{k}_\perp&=& \frac{p^+_2}{p^+_d}\vec{p}_{\perp}-\frac{p^+}{p^+_d}
\vec{p}_{2\perp} \label{eq:lf-k2} \,, \\
k^3&=&\frac{1}{2}(p^+-p^+_2)\sqrt{\frac{m^2+\vec{k}^{\,2}_\perp}{p^+p^+_2}} 
\label{eq:lf-k3} \,, \\
M_0&=&2\omega_m(\vec{k}) \label{eq:lf-k4} \,.
\end{eqnarray}
 
To use LF results in the calculation of DY cross section Eq.(\ref{eq:dy-pd}), we need to 
transform the LF variables into the usual variables. We thus have
\begin{eqnarray}
\rho_{{p}_d}(\vec{p}) &=&
(1+\frac{p^3}{\omega_m(\vec{p})}) \rho^{LF}_{\tilde{p}_d}(\tilde{p})
\end{eqnarray}

\section{Numerical Results }

\begin{figure}[t]
\begin{center}
\includegraphics[clip,width=0.8\textwidth]{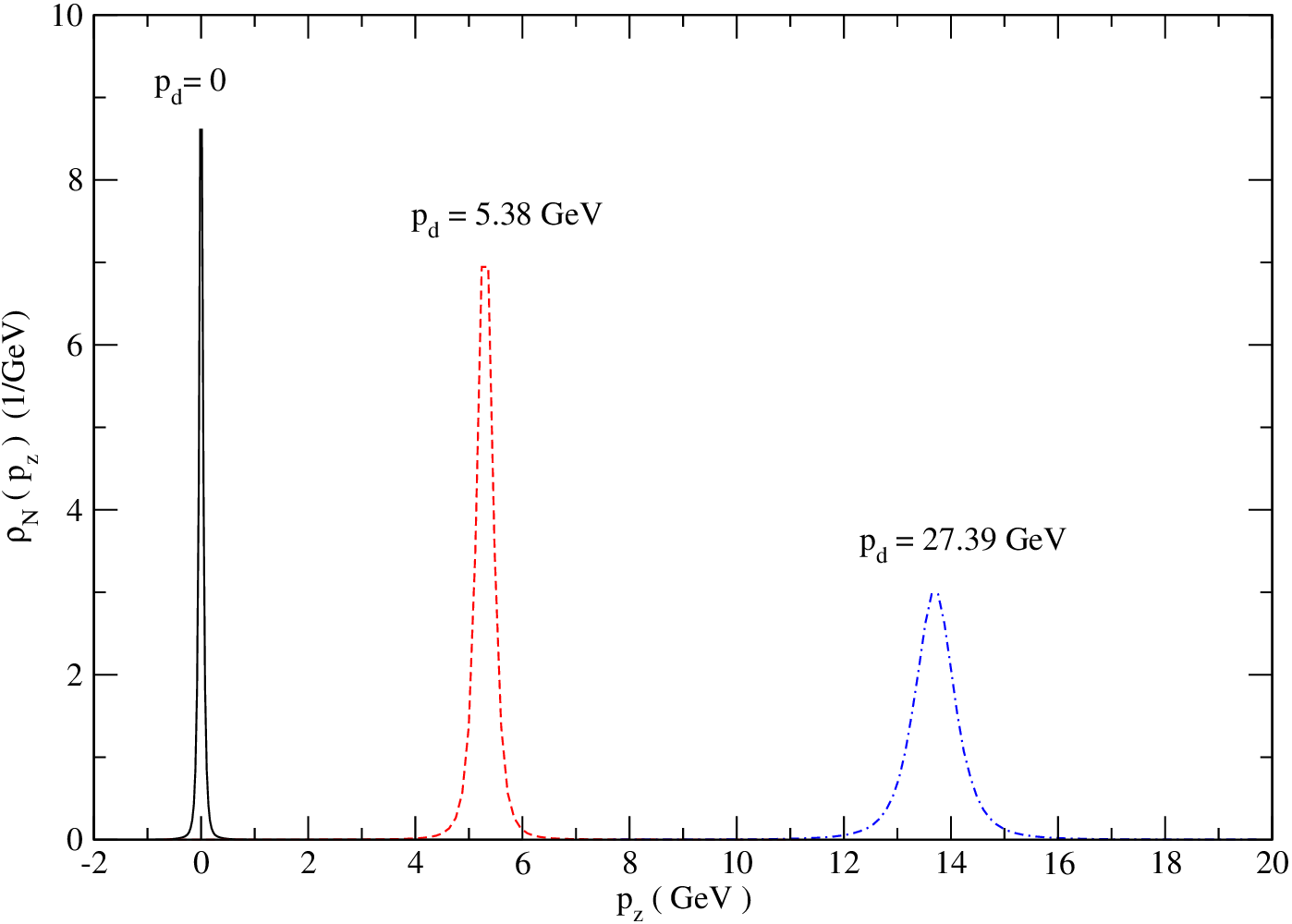}
\caption{Nucleon momentum distribution $\rho_{p_d}(p)$ of
 a deuteron moving with a momentum $p_d=|\vec{p}_d|$ in
$p$-$d$ collisions at the proton laboratory energies $E$= 0 (solid ), 
120 (dashed), 800 (dash-dotted) GeV}. 
 \label{fig:rho-v14}
\end{center}
\end{figure}

To calculate $p$+$d$ DY cross sections, 
we need to first determine the average nucleon momentum $p^z_{ave}$ in
Eq.(\ref{eq:pdf-ave}). This is done  by examining the dependence of
the nucleon momentum distribution $\rho_{p_d}(p)$ on the deuteron $p_d$ using
the formula given in subsection V.D.
In Fig.\ref{fig:rho-v14}, we show  $\rho_{p_d}(p_z)=\int \rho_{p_d}(\vec{p})dp_xdp_y $ 
of the Instant Form (IF) which are used in our calculations
for the $p$-$d$ DY cross sections  at $E=0, 120, 800$ GeV.
 The Argonne-V18 potential\cite{v18} is
used here to generate the deuteron wavefunction $\chi_d(\vec{k})$. 
 We see that as $p_d$ increases, the spreading of the
momentum distribution is wider. This is of course the consequence of
Lorentz contraction of the density distribution in coordinate space.
In the same figure, we also indicate the deuteron momentum $p_d$ for each case.
We observed that their peaks are near $p_d/2$ at each energy, and
we hence set $p^z_{ave} = p^z_d/2$ in the calculation of DY cross
sections using Eqs.(\ref{eq:pdf-ave}) and (\ref{eq:dy-pd-cc}).

We have also carried out calculations for the Light-Form (LF)  and find that the
results are indistinguishable from that shown in Fig.\ref{fig:rho-v14}.
This turns out to be due to the fact that in the
region where $|\chi(k)|^2$ dominant, the Jacobin 
$|\frac{\partial(\vec{p}_d,\vec{k})}{ \partial(\vec{p},\vec{p}_2)}|$ of IF
and
$|\frac{\partial(\tilde{p}_d,\vec{k})}{ \partial(\tilde{p},\tilde{p}_2)}|$ of LF
are almost the same numerically.

We use the CETEQ5m\cite{ceteq5m} parton distributions to perform calculations
using Eq.(\ref{eq:dy-pp-cc}) for $p$-$p$ and
Eq.(\ref{eq:pdf-ave})and Eq.(\ref{eq:dy-pd-cc}) for $p$-$d$
DY cross sections.
If we neglect the $p^z_N$-dependence in the
parton distribution function $f^q(x_2p^z_{ave}/p^z_N)$
in Eq.(\ref{eq:pdf-ave}) and set
$p^z_N \rightarrow p^z_{ave}$ such that
$f^q(x_2p^z_{ave}/p^z_N) \rightarrow f^q(x_2)$,
we then have
$\Gamma^{q}_{p_d,N}(x_2) \rightarrow f^q_N(x_2)$ since
$\int d\vec{p}_N \rho_{p_d}(\vec{p}_N)=1$ as defined by the normalization
of deuteron wavefunction $\chi_d$.
The calculations based on this simplification are then identical to
those from using Eqs.(\ref{eq:dy-exp})-(\ref{eq:dy-pd-exp})
of Ref.\cite{peng2001}. Clearly the results from using this
frozen-nucleon  approximation do not include the Fermi Motion (FM) effects.
Our interest is to see how these "No FM" results are
compared with the "With FM" results calculated from
using the convolution formula
Eqs.(\ref{eq:pdf-ave}) and (\ref{eq:dy-pd-cc}).

\begin{figure}[h]
\begin{center}
\includegraphics[clip,width=0.8\textwidth]{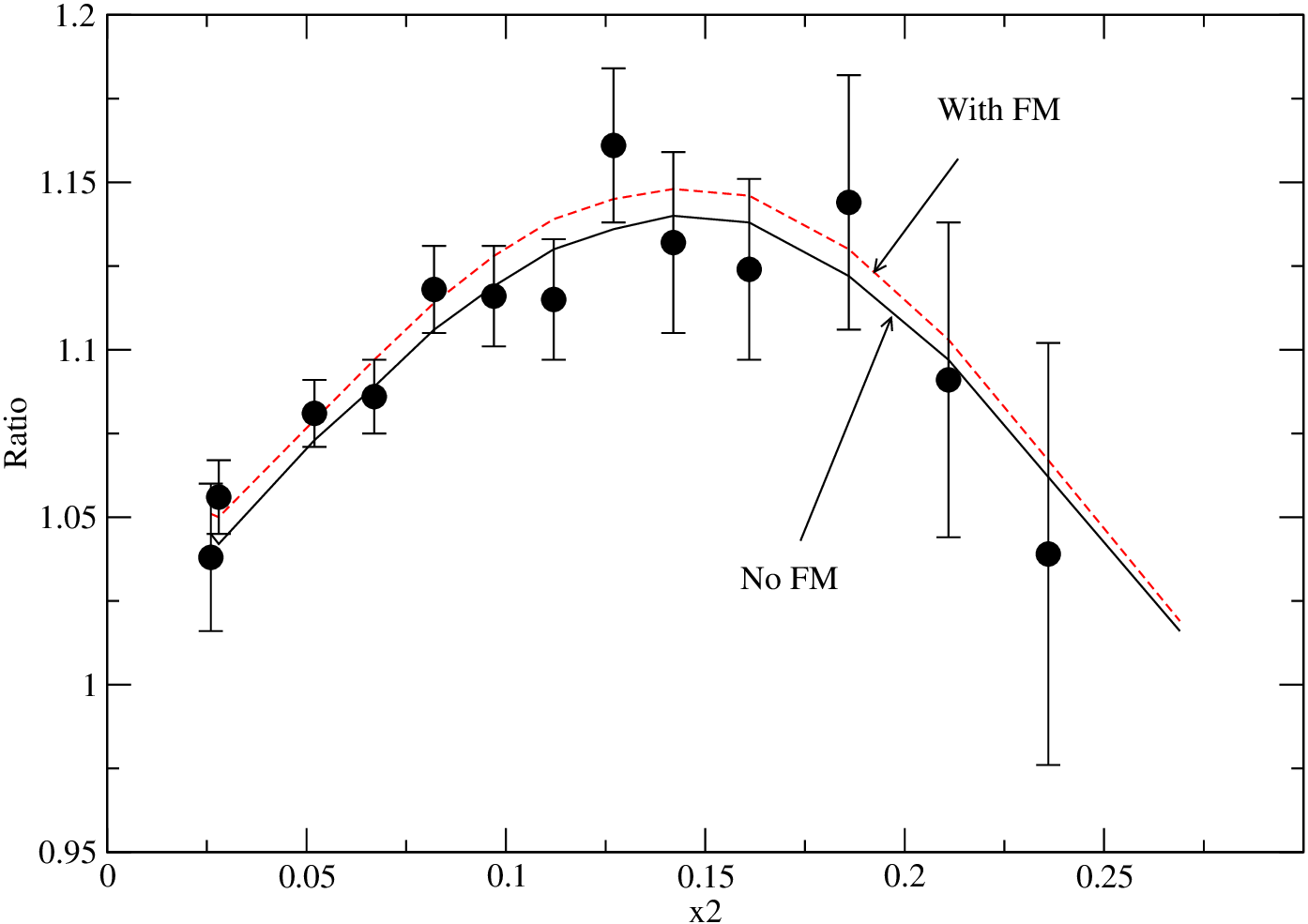}
\caption{Ratio $\sigma^{pd}/\sigma(pp)$ at $E=800 GeV$. The data are from
Ref.\cite{peng2001}.}
\label{fig:ratio-800}
\end{center}
\end{figure}

\begin{figure}[h]
\begin{center}
\includegraphics[clip,width=0.8\textwidth]{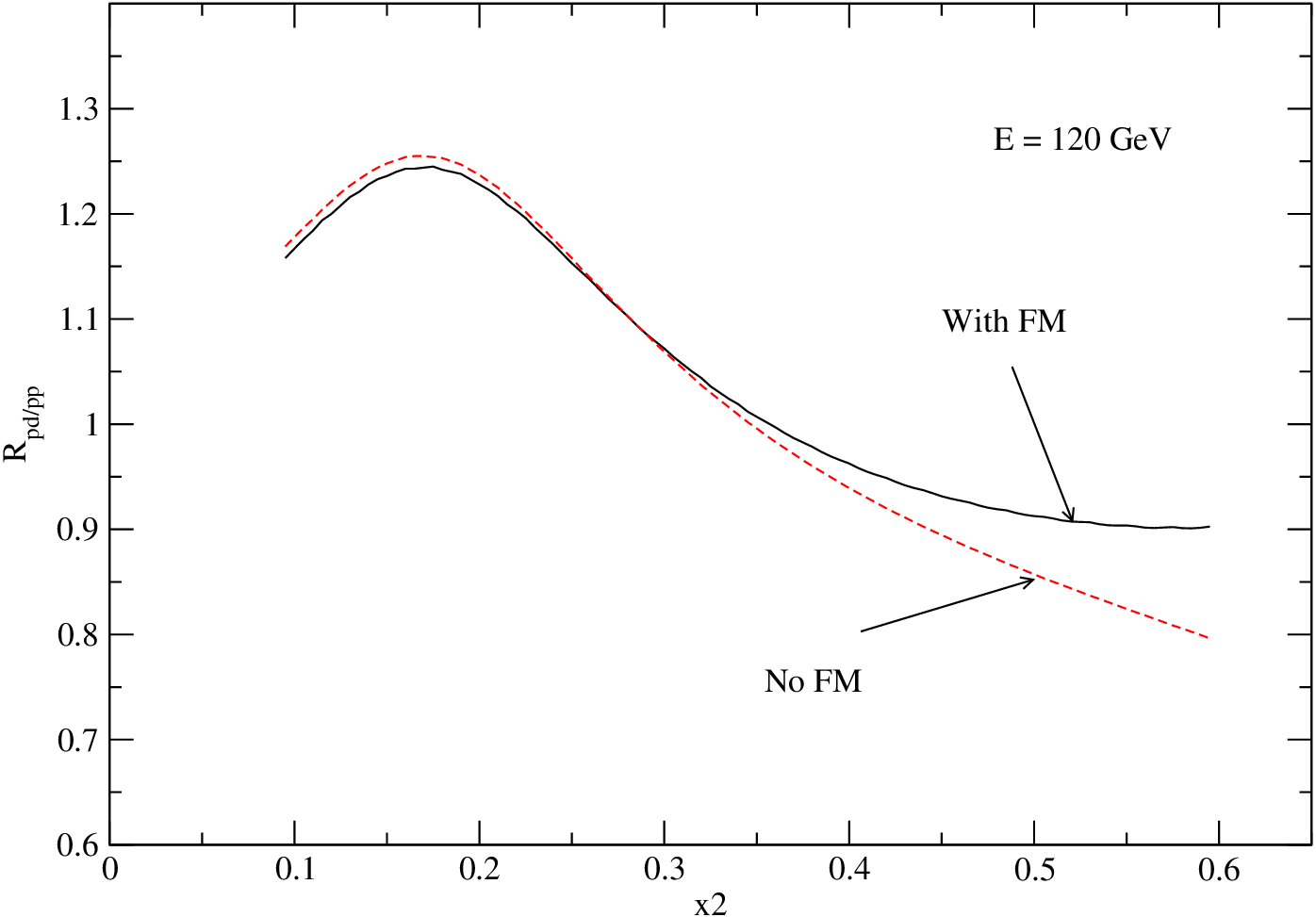}
\caption{Ratio $\sigma^{pd}/\sigma(pp)$ at $E=120 GeV$ and $x_1 = 0.6$. }
\label{fig:ratio-120-06}
\end{center}
\end{figure}

We first investigate the data at 800 GeV of Ref.\cite{peng2001}
by calculating the ratio
\begin{eqnarray}
R_{pd/pp}=\frac{d^{pd}(p,p_d)}{dx_1dx_2}/(2\times
\frac{d^{pp}(p,p_p)}{dx_1dx_2})\,.
\end{eqnarray}
Our results 
are shown in Fig.\ref{fig:ratio-800}.
We see that the Fermi motion
has less than $1\%$ effect on the ratio $\sigma^{pd}/(2\sigma^{pp})$
in the low $x_2 < $ 0.25 region covered by this experiment.
Our "No FM" results are similar to that presented
in Ref.\cite{peng2001} and agree with the data.

In Fig.\ref{fig:ratio-120-06}, we show the Fermi motion effect at 120 GeV 
which will be considered in the upcoming experiment at Fermi Laboratory.
The results are for  $x_1=0.6$.
We see that at small $x_2$, the FM effect is also very small as in 
 Fig.\ref{fig:ratio-800} for 800 GeV. However the FM effects become
increasingly larger with  $x_2$; about $25\%$ as $x_2\rightarrow 0.6$. 
This is similar to
what is known in the DIS studies.

\begin{figure}[h]
\begin{center}
\includegraphics[clip,width=0.8\textwidth]{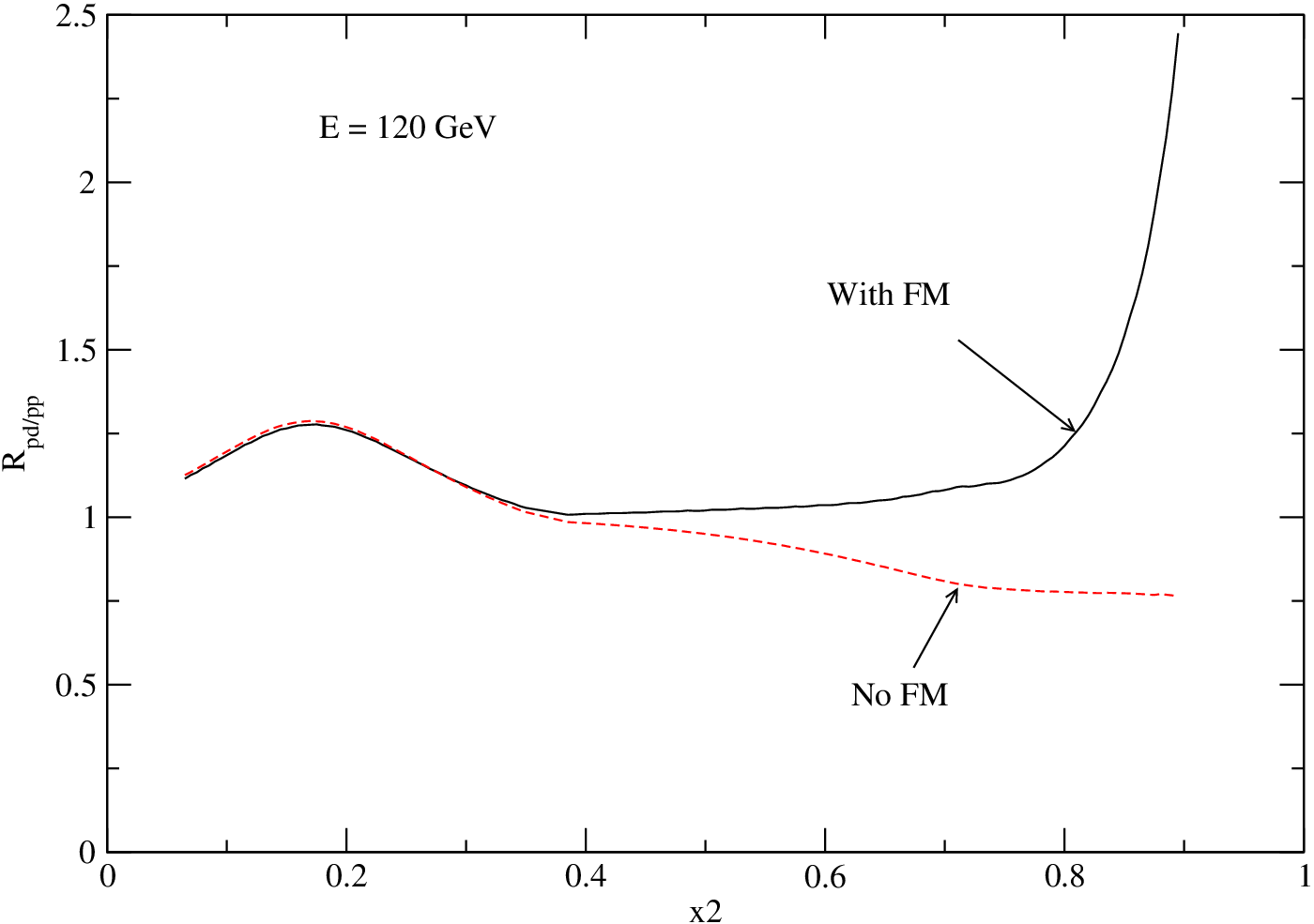}
\caption{Ratio $\sigma^{pd}/\sigma(pp)$ at $E=120 GeV$and $x_1=0.9$. }
\label{fig:ratio-120}
\end{center}
\end{figure}

\begin{figure}[h]
\begin{center}
\includegraphics[clip,width=0.8\textwidth]{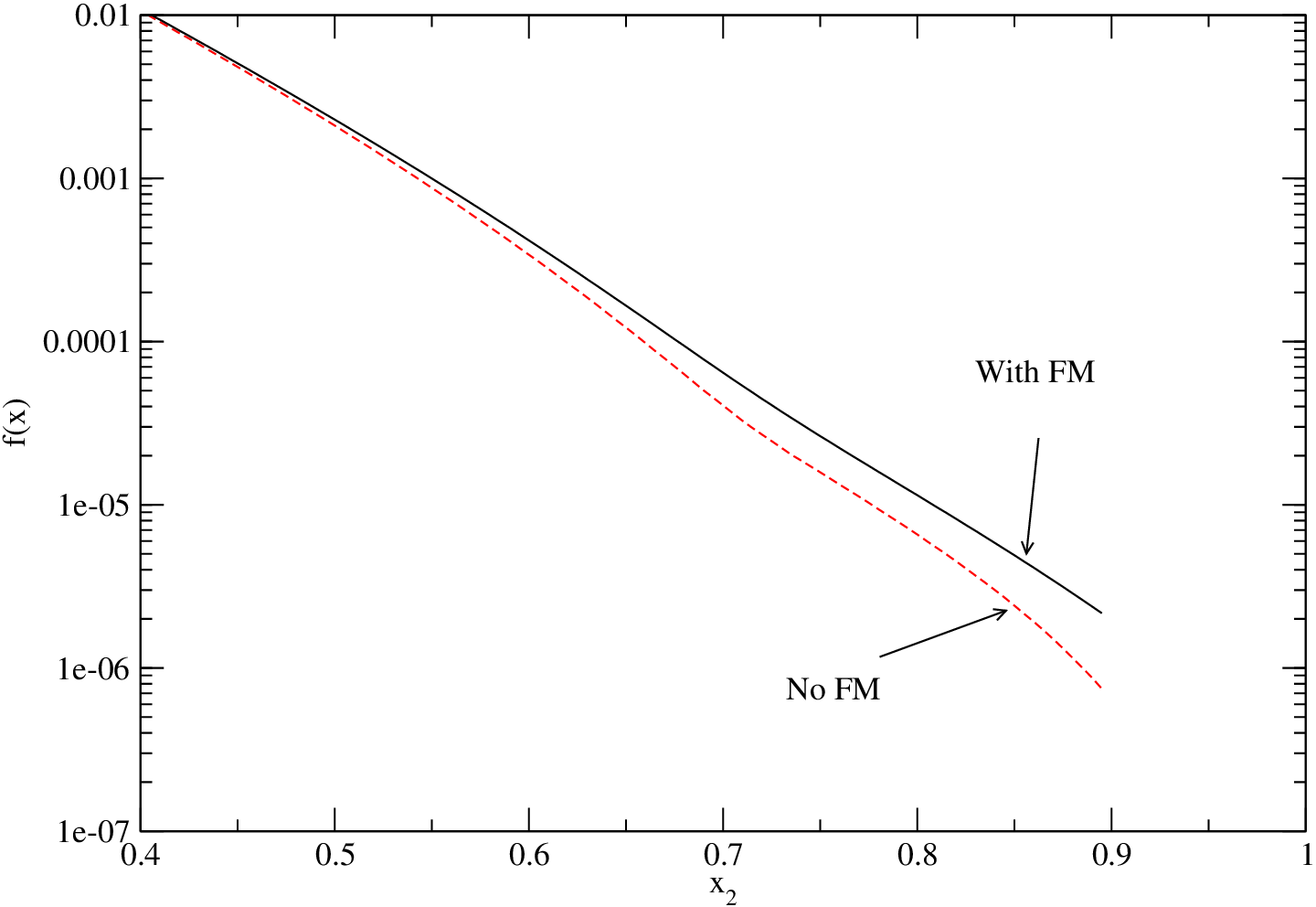}
\caption{Convoluted and factorized $\bar{u}$ distributions at $E=120 GeV$
and $x_1=0.9$.
The results for $\bar{d}$ are very similar and are therefore not shown. }
\label{fig:pdf-120-baru}
\end{center}
\end{figure}

To see the Fermi motion effects more clearly,
we present results for $x_1=0.9$ 
 in  Fig.\ref{fig:ratio-120}. We see that the ratio $\sigma^{pd}/(2\sigma^{pp})\rightarrow $
about 2.5 as $x_2 \rightarrow 0.9$.
This can be understood from Fig.\ref{fig:pdf-120-baru} where we show the
differences between the convoluted parton distribution $\Gamma^{\bar{u}}_{p_d,N}(x_2)$
, defined by Eqs.(\ref{eq:pdf-ave}), and $f^{\bar{u}}_N(x_2)$ of CETEQ5m.
We see that their differences can be as large as a factor of about 2 as
$x_2 \rightarrow 0.9$. Similar differences are also for other parton distributions.

The results presented in Figs.\ref{fig:rho-v14}-\ref{fig:pdf-120-baru} are from
using the deuteron wavefuncition $\chi(\vec{k})$ generated from Argonne-V18 NN potential.
Similar results have also been obtained by using the wavefunctions from Nijmegan\cite{nij}, CD-Bonn\cite{cdbonn},
and Paris\cite{paris} NN potentials.

\section{Summary and Discussions}
Within a formulation based on relativistic quantum mechanics, we
have developed convolution formula for calculating the $p$-$d$ 
Drell-Yan                            
cross sections using
the available parton distribution functions\cite{ceteq5m} and
the nucleon momentum distributions generated from realistic
 nucleon-nucleon potentials\cite{v18,nij,paris,cdbonn}.
When the Fermi motion in the deuteron is neglected,
our formula are reduced to the factorized form Eq.(2) which was
used in the previous analysis.

By comparing the results from the convoluted and factorized calculations, we
have shown that  in the small $x_2 < 0.3$ region,
the Fermi motion effect is very small and
our results for the ratios $R_{pd/pp}=\sigma^{pd}/(2\sigma^{pp})$
 agree well with the data at 800 GeV of Ref.\cite{peng2001}.
On the other hand, the Fermi motion effect can 
influence significantly the $p$-$d$ Drell-Yan cross sections
in the large
 Bjorken $x_2 > $ about 0.4 region.
At $120$ GeV 
the predicted Fermi Motion effect can enhance the ratios $R_{pd/pp}$ by
about 20 $\%$ at $x_2 \sim 0.6$ and about a factor of $2.5$ at
$x_2 \rightarrow 1.0$.
Our results suggest that the Fermi motion effect must be included in
extracting the $\bar{d}/\bar{u}$ ratio in the nucleon from the upcoming
experiment at Fermi Laboratory.

In this work, we have only 
investigated the most trivial nuclear effects
on DY processes. To extract the $\bar{d}/\bar{u}$ ratios precisely, 
we need to extend our formulation to
account for other nuclear effects. 
By using the approximations Eqs.(\ref{eq:delta-f}) and
(\ref{eq:app-2})  for the time-components of $\delta$-functions  
in our derivations, 
the binding effects on the partons in the nucleon and the undetected
fragments $X_p$ and $X_N$ are not treated rigorously.
To improve this incomplete treatment of binding effects, one
must depart from the simple convolution formula by performing
calculations using the spectral functions
of the nucleon in terms of parton degrees of freedom.
Such information is not available
at the present time. 

The use of impulse
approximation, as illustrated in Fig.\ref{fig:dy-mech-imp}
, only accounts for the leading-order term  of the
 full DY amplitude. It is also necessary to
investigate the accuracy of this approximation. For example, we perhaps
need to investigate the double scattering mechanisms which are considered
in Ref.\cite{mt93} to describe the shadowing effects on DIS.
We also need to investigate the effects due to pion excess, as 
discussed in Refs.\cite{miller84, koltun}.

Finally, we mention that there exists other relativistic formulations of
DIS and DY processes.
 The differences between different approaches are mainly
in describing the structure of the considered composite systems, nucleons or
nuclei, in terms of interactions between their constituents.
In the formulation considered in this work, all particles
are on their mass shells, but the systems can be off the energy shell
during interactions.
 Thus it is also necessary to examine how our formulation differs from
other formulations within which the
nucleons in the deuteron can be
off their mass-shell $p^2_N\neq m^2_N$, such as the
covariant formulation
given in Ref.\cite{mst94}. Similar theoretical questions also exist
in many-years investigations of nuclear dynamics at intermediate and
high energies; such as the differences\cite{kl74} between different nucleon-nucleon
potentials constructed using different three-dimensional reductions
of Bethe-Salpeter equations.

\begin{acknowledgments}
The author would like to thank D.F. Geesaman, R.J. Holt, and
J.-C. Peng for their helpful discussions.
This work is supported
by the U.S. Department of Energy, Office of Nuclear Physics Division, under
Contract No. DE-AC02-06CH11357.
\end{acknowledgments}

\end{document}